\documentclass{ieeeaccess}
\usepackage{cite}
\usepackage{amsmath,amssymb,amsfonts}
\usepackage{algorithmic}
\usepackage[usenames,dvipsnames]{xcolor}
\usepackage{pgfplots}
\usepackage{orcidlink}
\usetikzlibrary{shapes.geometric}
\usetikzlibrary{positioning}
\usetikzlibrary{quantikz}
\usepackage{romannum}
\usepackage{graphicx}
\usepackage[font={sf,scriptsize},           
            labelfont={bf,color=accessblue},
            caption=false]                  
           {subfig} 
\usetikzlibrary{arrows.meta,arrows}
\usepackage{circuitikz}
\usetikzlibrary{decorations.pathmorphing}
\usepackage{pgfplotstable}
\pgfplotsset{compat=1.7}
\usepackage{tikz}
 \NewSpotColorSpace{PANTONE}
  \AddSpotColor{PANTONE} {PANTONE3015C} {PANTONE\SpotSpace 3015\SpotSpace C} {1 0.3 0 0.2}
  \SetPageColorSpace{PANTONE}
\usepackage{graphicx}
\usepackage{textcomp}
\def\BibTeX{{\rm B\kern-.05em{\sc i\kern-.025em b}\kern-.08em
    T\kern-.1667em\lower.7ex\hbox{E}\kern-.125emX}}
\begin{document}
\history{Date of publication xxxx 00, 0000, date of current version xxxx 00, 0000.}
\doi{10.1109/ACCESS.2023.0322000}

\title{Microwave Gaussian quantum sensing with a CNOT gate receiver}
\author{\uppercase{Hany Khalifa}\authorrefmark{1,2} \orcidlink{0000-0002-1276-5428}, 
\uppercase{Kirill Petrovnin}\authorrefmark{2}, \uppercase{Riku J\"{a}ntti}\authorrefmark{1} \IEEEmembership{Senior, IEEE}, \uppercase{Gheorghe Sorin Paraoanu} \authorrefmark{2,3}}

\address[1]{Department of Information and Communications Engineering, Aalto University, Espoo, 02150, Finland }
\address[2]{QTF Centre of Excellence, Department of Applied Physics,
Aalto University, FI-00076 Aalto, Finland}
\address[3]{InstituteQ – the Finnish Quantum Institute, Aalto University, Finland}
\tfootnote{HK and RJ acknowledge funding from the Academy of Finland project Quantum Enhanced Microwave Backscatter Communications (grant number 319578). This work has received funding from the European Union’s Horizon 2020 research and innovation programme under grant agreement no. 862644 (FET-Open project: Quantum readout techniques and technologies QUARTET), supporting the work of KP and partly that of HK.  GSP acknowledges the project Business Finland QuTI (decision 41419/31/2020). In addition, the authors are grateful to Saab for the scientific collaboration under a research grant agreement with Aalto University. This work was performed as part of the Academy of Finland Centre of Excellence program (project 352925)}

\markboth
{Author \headeretal: Preparation of Papers for IEEE TRANSACTIONS and JOURNALS}
{Author \headeretal: Preparation of Papers for IEEE TRANSACTIONS and JOURNALS}

\corresp{Corresponding author: Hany Khalifa (e-mail: hany.khalifa@aalto.fi).}

\begin{abstract}
In \textit{quantum illumination} (QI) the non-classical correlations between \textit{continuous variable} (CV) entangled modes of radiation are exploited to detect the presence of a target embedded in thermal noise. The extreme environment where QI outperforms its optimal classical counterpart suggests that applications in the microwave domain would benefit the most from this new sensing paradigm. However all the proposed QI receivers rely on ideal photon counters or detectors, which are not currently feasible in the microwave domain. Here we propose a new QI receiver that utilises a CV \textit{controlled not gate} (CNOT) in order to perform a joint measurement on a target return and its retained twin. Unlike other QI receivers, the entire detection process is carried out by homodyne measurements and square-law detectors. The receiver exploits two squeezed ancillary modes as a part of the gate's operation. These extra resources are prepared offline and their overall gain is controlled passively by a single beamsplitter parameter. 
We compare our model to other QI receivers and demonstrate its operation regime where it outperforms others and achieves optimal performance. Although the main focus of this study is microwave quantum sensing applications, our proposed device can be built as well in the optical domain, thus rendering it as a new addition to the quantum sensing toolbox in a wider sense.
\end{abstract}


\titlepgskip=-21pt

\maketitle

\section{Introduction}
\label{sec:introduction}
\PARstart{Q}{uantum} sensing is a new paradigm for information detection that utilises non-classical features of the electromagnetic (EM) radiation to push detection sensitives beyond the classical limits. Entanglement, squeezing and superposition of quantum states are the main resources upon which many of the quantum sensing architectures are built \cite{RevModPhys.89.035002, pirandola2018advances}. Besides quantum sensing, quantum entanglement is an essential feature of many other quantum-technology applications, as it allows us to establish remote correlations between two jointly prepared EM radiation modes. Quantum cryptography \cite{PhysRevLett.67.661}, quantum computing \cite{nielsen2002quantum}, quantum communication \cite{gisin2007quantum} and quantum-enhanced metrology \cite{giovannetti2011advances}, have all exploited quantum entanglement to outperform their classical counterparts. Nonetheless, quantum entanglement is a fragile phenomenon, susceptible to environment-induced decoherence in the form of excess noise photons.

Quantum illumination is a quantum sensing protocol that can retrieve information sent over a noisy, entanglement-breaking channel \cite{lloyd2008enhanced}. The protocol utilizes entangled \textit{two mode squeezed vacuum} (TMSV) states to detect the presence or absence of a target embedded in a thermal environment \cite{tan2008quantum}. A probe (denoted as \textit{signal}) is sent to illuminate a target, while its twin (denoted as \textit{idler}) is retained in order to perform a correlation measurement on the target's return. The operational domain of QI is the low \textit{signal-to-noise ratio} (SNR) limit, where the optimum QI receiver enjoys a 6dB advantage in error exponent over the optimum classical one \cite{tan2008quantum}. This suggests that the microwave domain is probably the most natural setting for QI  experiments.  Unfortunately, up to this moment there is no known physical realization of the optimum QI receiver. Currently, up to 3 dB error exponent enhancement can be attained  theoretically with the available hardware. The \textit{optical parametric amplifier} (OPA), and \textit{phase conjugate} (PC) receivers \cite{guha2009gaussian} are the most remarkable receiver architectures that had been demonstrated experimentally to reap this sub-optimal advantage. The full 6 dB advantage can only be hypothetically attained with the complicated \textit{sum frequency generation} (SFM) receiver and its extremely intricate upgrade, \textit{feed-forward} SFG (FF-SFG) \cite{zhuang2017optimum}. Further, when non ideal storage of the idler mode is considered, the performance of all the mentioned receivers is greatly affected \cite{barzanjeh2015microwave}: 6 dB of idler loss is enough to rule out any quantum advantage.  However, all of the aforementioned designs relied on ideal photon counters operating in the low SNR regime in order to acknowledge a successful detection event. For quantum optical experiments, despite the insignificance of thermal background noise, efficient photon counting with low dark counts requires the use of superconductors and therefore operation at low temperatures. For microwave-frequency experiments, due to the extremely small powers at the single quantum level, microwave photon counters are only at the proof of concept stage.

In this article we propose a new microwave QI receiver that operates without the need for ideal single photon counters. Our proposed model is based on the CV CNOT operation \cite{furusawa2011quantum, filip2005measurement, yoshikawa2008demonstration, PhysRevA.71.055801}. Under this unitary gate the signal and idler quadratures transform into a superposition that is directly related to their cross-correlations features \cite{gu2009quantum, bruss2019quantum}.
Operationally speaking, a CV CNOT gate utilises two quadrature-squeezed ancillary modes, such that one is position-squeezed, while the other is momentum-squeezed. In order to avoid the cumbersome process of nonlinear coupling upon a receiving event, it has been demonstrated in \cite{filip2005measurement} that an offline preparation of the squeezed resources is both equivalent and more efficient than an online nonlinear coupling of a mode pair. The overall interaction gain can be controlled by a single beamsplitter parameter. 
This controlled operation gives us the ability to smoothly choose the operational domain where our device can outperform other QI receivers. 

\begin{figure}
\centering
    \begin{tikzpicture}
    \draw[thick] (0,0)--(0,0.5)--(1,0.5)--(1,-1)--(0,-1)--(0,0);
    \draw[](0.5,-0.25)--(0.5,-0.25) node [] {\scriptsize\text{TX}};
      \draw [-{Latex[length=2mm]},decorate,decoration={snake,amplitude=.4mm,segment length=2mm,post length=1mm}] (1,0) -- (2,0.6);
      \draw[] (1.5,0.5)--(1.5,0.5) node []{\scriptsize$a_{S}$};
        \node[cloud, draw=blue, minimum width = 0.3cm, minimum height = 0.3cm] (c) at (2.5,0.6) {?};
        \draw[](2.5,1.1)--(2.5,1.1) node[] {\scriptsize $\eta << 1$};
        \draw [-{Latex[length=2mm]},decorate,decoration={snake,amplitude=.4mm,segment length=2mm,post length=1mm}] (3,0.6) -- (4,0);
     \draw[](3.7, 0.4)--(3.7,0.4)node [] {\scriptsize $a_{R}$};
     \draw[thick] (4,0)--(4,0.5)--(5,0.5)--(5,-1)--(4,-1)--(4,0);
      \draw[](4.5,-0.25)--(4.5,-0.25) node [] {\scriptsize \text{RX}};
     \draw[thick] (1,-0.5)--(1.7,-0.5);
     \draw[thick, blue] (2.4,-0.5)--(2.6,-0.3);
     \draw[thick, blue] (2.4,-0.5)--(2.2,-0.7);
     \draw[-{Latex[length=2mm]},thick] (2,-0.5)--(4,-0.5);
      \draw[](1.4,-0.3)--(1.4,-0.3) node [] {\scriptsize $a_{I}$};
       \draw[](3.4,-0.3)--(3.4,-0.3) node [] {\scriptsize $a_{\rm M}$};
      \draw[thick, dotted] (2,-0.2)--(2.8,-0.2)--(2.8,-0.8)--(2,-0.8)--(2,-0.2);
      \draw[] (2.4,-1)--(2.4,-1) node []{\scriptsize\text{lossy memory}};
      \draw[] (2.35,-0.35)--(2.35,-0.35) node []{\scriptsize $T$};
     \draw[thick] (1.7,-0.5)--(2,-0.5);
      \draw [-{Latex[length=2mm]},decorate,decoration={snake,amplitude=.4mm,segment length=2mm,post length=1mm}] (4,1.5) -- (4,0.8);
      \draw[] (4,1.6)--(4,1.6) node[] {\scriptsize$\eta =0$};
       \draw[] (4.3,1.15)--(4.3,1.15) node[] {\scriptsize$a_{B}$};
    \end{tikzpicture}
    \caption{The quantum illumination protocol. At the transmitter correlated signal-idler pairs are generated, where the signal is sent towards a suspected region, while the idler is retained in a lossy memory element for a correlation measurement on the target's return. When there is no object, that is, $\eta=0$, the target return is a noisy environment mode $a_{B}$. }
    \label{fig:fig1}
    \end{figure}
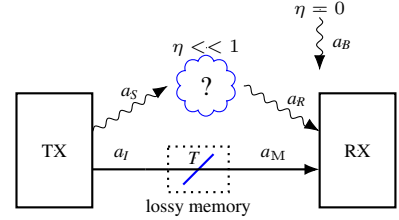

The basic idea behind the operation of the proposed receiver is simple: In the event of receiving a small fraction of the signal-idler initial correlations, the CNOT receiver strengthens it by a scalar value equal to the receiver's controllable interaction again. This is made possible due to the entangling properties of the universal CNOT gate. On the other hand, when these correlations are lost, the receiver outputs uncorrelated noise beams. Then the signal levels of both possible cases are determined by homodyning the receiver's output field quadratures \cite{haus2000electromagnetic, braunstein2005quantum}. However, since the average homodyne currents of the quadratures of a TMSV necessarily vanish, we propose feeding the output homodyne current to a square law detector, a spectrum analyzer (SA) for instance, in order to overcome this problem. This had been the standard method in the optical domain \cite{wu1986generation, furusawa2011quantum, furusawa2015quantum, ukai2014multi} and can be straightforwardly replicated in microwave quantum optics. Further, our device considers the non ideal storage of the idler mode Fig. (\ref{fig:fig1}), modelled by a beamsplitter with transmisivitty $T$, where the beamsplitter's unused port injects vacuum noise.
Finally it is worth mentioning that recently in the domain of microwave circuit quantum electrodynamics (CQED), there have been other successful implementations of the universal CNOT gate \cite{rosenblum2018cnot, PhysRevX.12.011008, krantz2019quantum}. 
We have opted for this specific implementation since its performance can be tracked analytically in a straightforward manner and can serve as a good model to calculate the receiver's internal noise. 
Thus, as long as a CNOT gate platform is capable of performing the gain-controlled, generalized CNOT interaction, one should be able to replicate the results of this study in both the microwave and optical domains.

This article is organized as follows: in Section \Romannum{2} we briefly describe the QI enhanced sensing protocol, then in section \Romannum{3} we describe the theory of operation of our CNOT receiver. We are mostly concerned with showing the ability of our receiver design to extract the signal-idler cross-correlations.  Section \Romannum{4} is mainly focused on the performance analysis of our device, a comparison between our model and OPA, PC and SFG receivers is carried out in great detail, such that the section's main objective is to demonstrate the operational regime where our device can outperform others. Finally, Section \Romannum{5} will be our conclusion.
\section{QI protocol}
We consider applications where a transmitter is sending its information over a noisy and lossy channel. The receiver, potentially co-located with the transmitter, stores a mode that shares quantum correlations with the transmitted signal. Quantum target detection \cite{shapiro2009quantum}, quantum radars \cite{lanzagorta2011quantum} and quantum backscatter communications\cite{jantti2020quantum} are perfect examples of these applications.

At the transmitter a pump field excites a non-linear element to generate $K$ independent signal-idler mode pairs via \textit{spontaneous parametric down conversion} (SPDC), $\{a^{j}_{S}, a^{j}_{I}\}, \hspace*{0.2 em} 0 \leq j \leq M$ \cite{esposito2022observation, perelshtein2022broadband, petrovnin2023generation}. The total number of probe signals is equal to  $K =  \tau W $, where $\tau$ is  
the duration of a transmission event and $W$ is the phase-matching bandwidth of the nonlinear element. For our purposes the archetypal nonlinear element in the microwave domain is the \textit{Josephson parametric amplifier} (JPA). Depending on the design, the operating frequency of a JPA is in the range of $4-8 \text{GHz}$ \cite{petrovnin2023generation}. Another celebrated device that can generate microwave signal-idler entangled pairs is the \textit{Josephson ring modulator} (JRM) \cite{PhysRevLett.109.183901, abdo2013nondegenerate}. In the case of a JPA source, the bandwidth of the generated twin pairs is typically $1 \text{MHz}$, hence for a total number of probe pairs $K=10^{6}$, the protocol duration would be $\tau \approx 1s$. However, the bandwidth can be substantially increased up to $100\text{MHz}$ when a \textit{travelling wave parametric amplifier} (TWPA) \cite{perelshtein2022broadband} source is utilized. This would dramatically result in a faster QI protocol.

Each signal-idler pair is in a TMSV that admits a number state representation,
\begin{align}
    \lvert \Psi \rangle_{SI} = \underset{n=0}{\overset{\infty}{\sum}} \sqrt{\frac{N_{S}^{n}}{(N_{S}+1)^{n+1}}} \lvert {n} \rangle_{S} \lvert {n} \rangle_{I},
\end{align}
where $N_{S}$ is the mean photon number in each of the signal and idler modes, i.e., $\langle a^{\dagger}_{S}a_{S} \rangle=\langle a^{\dagger}_{I}a_{I}\rangle =N_{S}$.

It is also useful to express the above TMSV as a squeezing operation applied to a vacuum state,
\begin{align}
    &\lvert \Psi \rangle _{SI} = \mathcal{S}(\gamma)\lvert  0,0\rangle_{SI},
    & S(\gamma)=e^{(\gamma a^{\dagger}_{S}a^{\dagger}_{I}-\gamma^{*}a_{S}a_{I})},
\end{align}
where the complex squeezing parameter $r$ is defined as $\gamma = r e^{i \varphi}$, and $\varphi$ is the angle of the squeezing axis. Relating the two expressions to each other, the number of photons in either the signal or idler mode can be redefined in terms of the complex squeezing parameter as $\langle a^{\dagger}_{S}a_{S} \rangle = \langle a^{\dagger}_{I}a_{I} \rangle =\sinh^{2}{r}$.

The entanglement between each pair is quantified by their $4 \times 4$ covariance matrix. 
\begin{align}
    C( S, I)= \begin{pmatrix}
        A & B \\
        B^\top & D
    \end{pmatrix},
\end{align}
where the matrices $A, B, B^{\top}, D$ are defined as follows:
$ A_{kl} =(1/2) [\langle q^{k}_{S}q^{l}_{S}+q^{l}_{S}q^{k}_{S} \rangle-\langle q^{k}_{S}\rangle \langle q^{l}_{S} \rangle ] = \text{diag}(N_{S}+1/2, N_{S}+1/2) $, $B_{kl} = (1/2)[\langle q^{k}_{S}q^{l}_{I}+q^{l}_{S}q^{k}_{I} \rangle -\langle q^{k}_{S} \rangle \langle q^{l}_{I}\rangle ]= \text{diag}([N_{S}(N_{S}+1)]^{1/2}, [N_{S}(N_{S}+1)]^{1/2})$, $ B^\top _{kl} = (1/2)[\langle q^{k}_{I}q^{l}_{S}+q^{l}_{I}q^{k}_{S} \rangle -\langle q^{k}_{I} \rangle \langle q^{l}_{S} \rangle ] = \text{diag}([N_{S}(N_{S}+1)]^{1/2}, [N_{S}(1+N_{S})]^{1/2})$, $D_{kl} = (1/2)[\langle q^{k}_{I}q^{l}_{I}+q^{l}_{I}q^{k}_{I} \rangle-\langle q^{k}_{I} \rangle \langle q^{l}_{I}\rangle ]  = \text{diag}(N_{S}+1/2, N_{S}+1/2)$, $ k,l =1,2$, $q^{1}= X = (a+a^{\dagger})/\sqrt{2}$, $ q^{2} = Y = -i(a-a^{\dagger})/\sqrt{2}$ are the mode quadratures, $B^\top$ is the transpose of $B$ and \text{diag} is a $2 \times 2$ diagonal matrix.

As for channel considerations, we will assume a lossy transmission medium overwhelmed by noise photons. Hypothetically, two transmission scenarios might arise under this channel model.

\textit{Hypothesis 1} (alternative hypothesis)$(H=1)$:
The transmitted signal reaches the receiver with a very small probability. This can be modelled by a low transmissivity beamsplitter that mixes the signal with a bath mode 
\begin{align}
    a_{R} = \sqrt{\eta} a_{S} + \sqrt{1-\eta} a_{B} ,
\end{align}
where $ a_{R}$ is the received mode, $\eta$ is the beamsplitter's transmissivity, and $a_{B}$ is a zero mean Langevin bath mode, $\langle a_{B} \rangle=\langle a_{B}^{\dagger} \rangle=0$, with mean photon number $\langle a_{B}^{\dagger}a_{B} \rangle = N_{B}$, 
and zero cross correlations $\langle a_{B_{k}}a_{B_{l}} \rangle=0, \forall k\neq l$, since a thermal state is diagonal in the number basis. 

\textit{Hypothesis 0} (null hypothesis) $(H=0)$: The transmitted signal is completely lost and replaced by a bath mode, $a_{B}$, i.e, $a_{R}=a_{B}$.

Simultaneously, we consider storing the idler mode in a leaky memory element, which can be represented by a pure loss channel with transmisivitty $T$
\begin{align}
    a_{\mathrm M} = \sqrt{T}a_{I}+\sqrt{1-T}a_{V}
\end{align}
where $a_{V}$ is an environment vacuum mode. It is worth noting that recently there has been some notable progress regarding microwave quantum memories, with efficiency as high as $80\%$ \cite{flurin2015superconducting, moiseev2018broadband}.
\section{Theory of operation}
\subsection{The CNOT receiver}
The preferred operating regime where QI's advantage is manifest is the low SNR, such that, $N_{S} \ll 1 \ll N_{B}$. In this domain the signal-idler cross correlations of a TMSV, $\langle a_{S}a_{I} \rangle + \langle a^{\dagger}_{S}a^{\dagger}_{I} \rangle$, where $\langle a_{S}a_{I} \rangle=(1/2) \big \langle (X_{S}+iY_{S})(X_{I}+iY_{I})\big\rangle = \big \langle X_{S} X_{I} \big \rangle - \big \langle Y_{S}Y_{I}\big\rangle =  \sqrt{N_{S}(1+N_{S})}$, $\langle a^{\dagger}_{S}a^{\dagger}_{I} \rangle =(1/2) \big \langle (X_{S}-iY_{S})(X_{I}-iY_{I})\big\rangle = \big \langle X_{S} X_{I} \big \rangle - \big \langle Y_{S}Y_{I}\big\rangle =  \sqrt{N_{S}(1+N_{S})}$, $\langle a_{S}a_{I} \rangle + \langle a^{\dagger}_{S}a^{\dagger}_{I} \rangle = 2\sqrt{N_{S}(1+N_{S})}$, exceed the maximum that can be attained classically with equal strength, uncorrelated coherent pairs, with average photon number $N_{S}$ each, $\langle \alpha \lvert X_{\mathrm m} \lvert \alpha \rangle  =\sqrt{2} \hspace*{0.2em}\text{Re}(\alpha) $, $\langle \alpha \lvert Y_{\mathrm m} \lvert \alpha \rangle= \sqrt{2} \hspace*{0.2em}\text{Im}(\alpha)$, $\langle \alpha,\alpha \lvert X_{\mathrm m}X_{\mathrm m} \lvert \alpha,\alpha \rangle = 2 \hspace*{0.2em}[\text{Re}(\alpha)]^{2} $, $ \langle \alpha,\alpha \lvert Y_{\mathrm m}Y_{\mathrm m} \lvert \alpha,\alpha \rangle  = 2 \hspace*{0.2em}[\text{Im}(\alpha)]^{2}$, $\langle \alpha,\alpha \lvert X_{\mathrm m}Y_{\mathrm m} \lvert \alpha,\alpha \rangle  = 2 \lvert \alpha \lvert^{2} =2 N_{S}$, where $\mathrm m$ $= S,I$, $\alpha$ is the coherent field complex amplitude, and $[X_{\mathrm m}, Y_{\mathrm n}] =i \delta_{\mathrm m \mathrm n}$ is the field quadratures commutation relation, where the reduced Planck's constant is set equal to one, $\hbar = 1$. 

A receiver's main task in QI is to extract the aforementioned signal-idler cross correlations from a target return and its retained idler twin, $\langle a_{R}a_{\mathrm M} \rangle$. In terms of the quadrature operators, this quantity can be expanded into four terms $(1/2) \big[ X_{R}X_{\mathrm M}+iX_{R}Y_{\mathrm M}+iY_{R}X_{\mathrm M}-Y_{R}Y_{\mathrm M}\big]$. Attempting to perform a \textit{heterodyne} measurement on each mode separately is probably the most economic way to gain full access to the field's quadratures \cite{yuen1980optical}. However, the splitting of  each mode first on a balanced beamsplitter, would add an additional 3dB loss to the overall output SNR \cite{schumaker1984noise, haus2000electromagnetic}. Along with detectors inefficiencies, this would rule out any quantum advantage over the optimal classical illumination (CI) receiver. Further, the Gaussian Wigner statistics of a directly homodyned squeezed state is non-negative and a nonlinear detection scheme, such as photon counting, is needed to reveal the non classical signal-idler signature \cite{knill2001scheme, karsa2022noiseless, shapiro2020quantum}. In this regard, the previous installments of the QI protocol opted for single photon counters as detectors for their receiver designs. 

In order to avoid the extra losses of double homodyning, and the status-quo technological infeasibility of microwave single photon counters, our proposed CV CNOT receiver mediates a controllable interaction between a target's return and a stored idler that can access their non-classical correlations by creating an observable quantity corresponding to their relative momentum and total position quadratures. Further, as mentioned in the introduction, the controllable gain of the receiver can in fact strengthen these correlations rendering them more visible for successful detection. Finally, in order to satisfy the required detector's non linearity described previously, our receiver utilises a square law detection chain \cite{barzanjeh2020microwave}, composed of a balanced double port homodyne detector and a spectrum analyzer, where the corresponding measurement outcome is the quadratures variances or powers \cite{ukai2014multi}.  This has been the standard method of detection in quantum optical experiments \cite{wu1986generation, furusawa2015quantum, furusawa2011quantum, haus2000electromagnetic}. For completeness we expose the details of this method in appendix \ref{sec:Analysis}. Besides making the above arguments more rigorous,  we now focus on the mathematical representation of our proposed device. We first show how it extracts the signal-idler cross correlation signature, then we see how the device's controllable gain can enhance it.

The CNOT receiver transforms a returned mode and its stored idler twin as follows,
\begin{align}
    X^{(\text{out})}_{R } &= e^{i G Y_{\mathrm M} X_{R}} \hspace*{0.2em} X_{R} \hspace*{0.2em} e^{-i G Y_{\mathrm M} X_{R}} \nonumber \\
    &= X_{R} \nonumber \\
      Y^{(\text{out})}_{R}  &=e^{i G Y_{\mathrm M} X_{R}} \hspace*{0.2em} Y_{R} \hspace*{0.2em} e^{-i G Y_{\mathrm M} X_{R}} \nonumber \\
      &=Y_{R} + [i G X_{R} Y_{\mathrm M},Y_{R}]+0, \nonumber \\
   &= Y_{R}- G Y_{\mathrm M} \nonumber\\
     X^{( \text{out})}_{\mathrm M} &= e^{i G Y_{\mathrm M} X_{R}} \hspace*{0.2em} X_{\mathrm M} \hspace*{0.2em} e^{-i G Y_{\mathrm M} X_{R}} \nonumber \\
     &=X_{\mathrm M} + [i G X_{R} Y_{\mathrm M},X_{\mathrm M}]+0, \nonumber \\
   &= X_{\mathrm M}+GX_{R}, \nonumber \\
    Y^{(\text{out})}_{\mathrm M} &= e^{i G Y_{\mathrm M} X_{R}} \hspace*{0.2em} Y_{\mathrm M} \hspace*{0.2em} e^{-i G Y_{\mathrm M} X_{R}}\nonumber \\
     &=Y_{\mathrm M} 
     \label{eq:CNOTRXOUT}
\end{align} 
where we have utilised the operator expansion formula for any two non commuting operators $[A,B] \neq 0$, $e^{\lambda A}Be^{-\lambda A}=B+\lambda [A,B]+\big(\lambda^{2}/2!\big) [A,[A,B]]+..$.., the commutation relation between the field's quadratures $[X, Y]=i$, such that $\hbar =1$, and $G$ is the interaction gain \cite{yoshikawa2008demonstration}. Note that  $[X^{(\text{out})}_{R },  Y^{(\text{out})}_{R }] = [X^{(\text{out})}_{M },  Y^{(\text{out})}_{M }] = i$ and the rest of the commutators are zero, as expected for a unitary transformation.

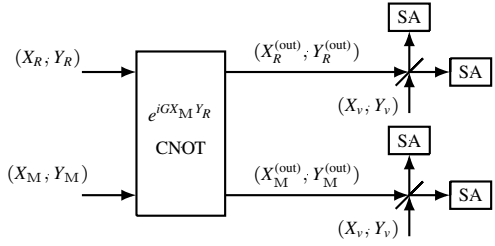
\begin{figure}
    \centering
    \begin{tikzpicture}[scale=0.9]
    \draw [thick] (-0.2,-0.2)--(1.1,-0.2)--(1.1,2.2)--(-0.2,2.2)--(-0.2,-0.2);
   
     \draw[-{Latex[length=2mm]},thick] (-1,0.1)--(-0.2,0.1);
 
    \draw[-{Latex[length=2mm]},thick] (-1,1.9)--(-0.2,1.9);

    \draw[-{Latex[length=2mm]},thick] (1.1,0.1)--(3.8,0.1);
 \draw[thick](3.8,0.1)--(4,0.3)--(3.6,-0.1);
    \draw[-{Latex[length=2mm]},thick] (1.1,1.9)--(3.8,1.9);
    \draw[thick](3.8,1.9)--(4,2.1)--(3.6,1.7);
 \draw[thick, -{Latex[length=2mm]}](3.8,-0.5)--(3.8,0.1);
     \draw[thick, -{Latex[length=2mm]}](3.8,0.1)--(3.8,0.7);
    \draw[thick](4.4,1.9)--(4.4,2.1)--(4.4,1.7)--(5,1.7)--(5,2.1)--(4.4,2.1);
     \draw[thick](3.8,0.7)--(4.1,0.7)--(3.5,0.7)--(3.5,1.1)--(4.1,1.1)--(4.1,0.7); 
\draw[thick, -{Latex[length=2mm]}](3.8,0.1)--(4.4,0.1);

    \draw[thick, -{Latex[length=2mm]}](3.8,1.3)--(3.8,1.9);
     \draw[thick, -{Latex[length=2mm]}](3.8,1.9)--(3.8,2.5);
     \draw[thick, -{Latex[length=2mm]}](3.8,1.9)--(4.4,1.9);
     \draw[thick](4.4,0.1)--(4.4,0.3)--(4.4,-0.1)--(5,-0.1)--(5,0.3)--(4.4,0.3);
     \draw[thick](3.8,2.5)--(4.1,2.5)--(3.5,2.5)--(3.5,2.9)--(4.1,2.9)--(4.1,2.5);
     \draw[](4.7,1.9)--(4.7,1.9) node [] {\scriptsize \text{SA}};
        \draw[](4.7,0.1)--(4.7,0.1) node [] {\scriptsize \text{SA}};
           \draw[](3.8,0.9)--(3.8,0.9) node [] {\scriptsize \text{SA}};
            \draw[](3.8,2.7)--(3.8,2.7) node [] {\scriptsize \text{SA}};
             \draw[](3.2,-0.4)--(3.2,-0.4) node [] {\scriptsize $(X_{v},Y_{v})$};
             \draw[](3.2,1.4)--(3.2,1.4) node [] {\scriptsize $(X_{v},Y_{v})$};
    \draw[](0.45,1.3)--(0.45,1.3) node []{\scriptsize $e^{i G X_{\mathrm M} Y_{R}}$};
     \draw[](0.45,0.8)--(0.45,0.8) node []{\scriptsize $\text{CNOT}$};
    \draw[](-1.5,0.4)--(-1.5,0.4) node [] {\scriptsize $(X_{\mathrm M},Y_{\mathrm M})$};
 ;
    \draw[](-1.5,2.1)--(-1.5,2.1) node [] {\scriptsize $(X_{R},Y_{R})$};

 \draw[](2.3,0.4)--(2.3,0.4) node [] {\scriptsize $(X^{\text{(out)}}_{\mathrm M},Y^{\text{(out)}}_{\mathrm M})$};
   
    \draw[](2.3,2.2)--(2.3,2.2) node [] {\scriptsize $(X^{\text{(out)}}_{R}, Y^{\text{(out)}}_{R})$};
\end{tikzpicture}
    \caption{A unitary gate representation of the CNOT receiver. Each spectrum analyzer (SA) module comprises a double port homodyne detector as described in appendix \ref{sec:Analysis}. The tuple $(X_{v},Y_{v})$ represents a vacuum mode. }
    \label{fig:Unitary}
\end{figure}
In Fig. (\ref{fig:Unitary}) we depict the CNOT receiver as a unitary gate. As can be seen, the receiver has two different quadrature input tuples, of two elements each, and their corresponding two observable output tuples. The first element of the first output tuple, $X^{\text{(out)}}_{R}$, is the unaffected mode, whereas the second, $Y^{\text{(out)}}_{R}$, carries information on both the returned and stored momentum quadratures. On the other hand, the first element of the second output tuple, $X^{\text{out}}_{\mathrm M}$, carries the position information of both the return and stored modes, whereas the second, $Y^{(\text{out})}_{\mathrm M}$, is the unaffected mode. Ideally this sort of interaction is probed in order to perform a \textit{non demolition} measurement on the unaffected quadratures by only measuring the translated ones \cite{caves1980measurement, braunstein2005quantum}. After that each output is mixed on a balanced beamsplitter with a vacuum mode, defined by its corresponding conjugate quadratures tuple, $(X_{v},Y_{v}) $. Finally, the powers of the four outputs are measured by a spectrum analyzer module (SA) comprising a double port homodyne detector (see details in appendix \ref{sec:Analysis}). In order to verify a successful implementation of the receiver's operation, and hence a successful capturing of sought cross correlations,  the four conjugate quadratures corresponding to the output return and memory modes have to be measured simultaneously. As can be deduced from Eq. (\ref{eq:CNOTRXOUT}), the power (second moment) of the unaffected quadrature is added to the translated one respectively for each output. In the event of a failed operation, that is, $G=0$, the powers are equal respectively.   
We now proceed with calculating the mode variances, i.e., (signal powers), of the involved quadratures as being measured practically, and demonstrate the previous ideas mathematically.

\subsection{Extracting the signal-idler cross correlation}
The receiver's outputs as demonstrated in Fig. (\ref{fig:Unitary}) are mode quadratures. The information contained in the signal-idler cross correlations can be accessed by measuring their respective variances, which in the present case coincides with the signal power (second moment), since we are dealing with zero mean fields. As pointed out earlier, in order to measure the quadratures variances simultaneously from the signal return and stored idler, $a^{\text{(out)}}_{R}$, $a^{\text{(out)}}_{\mathrm M}$, the modes are first split individually on a balanced beamsplitter, where a vaccuum mode enters from the unused port. Then detected by a square law detector. This would result in a 3dB loss of the measured quadrature. For illustration let's consider the first receiver's output, after a balanced beamsplitter the first output is homodyned for the position quadrature, thus $\bar{a}^{\text{(out)}}_{R} = (1/\sqrt{2}) (a^{\text{(out)}}_{R}+a_{v})$, $\bar{X}^{\text{(out)}}_{R} = (1/2)(a^{\text{(out)}}_{R}+a_{v}+ {a^{\text{(out)}}_{R}}^{\dagger}+a^{\dagger}_{v} )$ ,$\big \langle [\Bar{X}^{\text{(out)}}_{R}]^{2} \big \rangle = (1/2) \big \langle (X^{\text{(out)}}_{R}+X_{v})(X^{\text{(out)}}_{R}+X_{v}) \big \rangle =(1/2) \big (\big \langle [X^{\text{(out)}}_{R}]^{2} \big \rangle + \big \langle X^{2}_{v} \big \rangle \big )$. Similarly the second beamsplitter output can be homodyned for the momentum quadrature $Y^{\text{(out)}}_{R}$. As can be seen the 3dB noise penalty is visible now in the signal's power as the intensity of the original field is halved. The rest of the quadratures corresponding to the receiver's second output are treated in a similar manner. Thus, to keep the notations simple, we include the noise penalty directly to the following calculations, whereas the overall vacuum noise is added at the end of this derivation.  

Suppose that the alternative hypothesis is true, i.e, $H=1$, then
\begin{align}
    \langle [X^{(\text{out})}_{\mathrm M}]^{2} \rangle = \langle [X^{2}_{\mathrm M}+2GX_{\mathrm M}X_{R}+G^{2}X^{2}_{R}] \rangle, 
    \label{eq:CNOTOUT1}
\end{align}
where, 
\begin{align}
\langle X^{2}_{\mathrm M} \rangle &= \frac{1}{4}\big \langle (\sqrt{T} a_{I}+\sqrt{1-T}a_{V}+\sqrt{T}a^{\dagger}_{I}+\sqrt{1-T}a^{\dagger}_{V})  \nonumber \\ 
&\hspace*{0.5em}(\sqrt{T} a_{I}+\sqrt{1-T}a_{V}+\sqrt{T}a^{\dagger}_{I}+\sqrt{1-T}a^{\dagger}_{V}) \big \rangle \nonumber \\
&=\frac{(2TN_{S}+1)}{4} \nonumber \\
\langle X_{\mathrm M}X_{R} \rangle &= \frac{1}{4} \big \langle (\sqrt{T} a_{I}+\sqrt{1-T}a_{V}+\sqrt{T}a^{\dagger}_{I}+\sqrt{1-T}a^{\dagger}_{V}) \nonumber \\ 
 &\hspace*{1em} (\sqrt{\eta} a_{S}+\sqrt{1-\eta}a_{B} +\sqrt{\eta}a^{\dagger}_{S}+\sqrt{1-\eta}a^{\dagger}_{B})\big \rangle \nonumber \\
 &= \frac{\sqrt{\eta T N_{S}(1+N_{S})}}{2}  \nonumber \\
    \langle X^{2}_{R} \rangle &= \frac{1}{4} \big \langle (\sqrt{\eta} a_{S}+\sqrt{1-\eta}a_{B}+\sqrt{\eta}a^{\dagger}_{S}+\sqrt{1-\eta}a^{\dagger}_{B}) \nonumber \\
    &\hspace*{1em}(\sqrt{\eta} a_{S}+\sqrt{1-\eta}a_{B}+\sqrt{\eta}a^{\dagger}_{S}+\sqrt{1-\eta}a^{\dagger}_{B}) \big \rangle \nonumber \\ 
&=\frac{1}{4} \big(  \eta [1 +2\langle a^{\dagger}_{S}a_{S}\rangle]+(1-\eta)[1 +2 \langle a^{\dagger}_{B}a_{B} \rangle]\big) \nonumber \\
&=\frac{\big(  \eta(1+ 2N_{S}) +(1-\eta)(1+2N_{B})\big)}{4} 
\label{eq:derv1}
\end{align}
where in Eq. (\ref{eq:derv1}) we have used the following, $\langle a_{S}a_{S}\rangle = \langle a_{I}a_{I}\rangle = \langle a_{B} a_{B} \rangle = \langle a^{\dagger}_{S}a^{\dagger}_{S} \rangle =\langle a^{\dagger}_{I}a^{\dagger}_{I} \rangle = \langle a^{\dagger}_{B}a^{\dagger}_{B} \rangle= \langle a^{\dagger}_{S}a^{}_{I} \rangle = \langle a^{}_{S}a^{\dagger}_{I} \rangle = 0$, $\langle a_{S}a_{B}\rangle = \langle a_{I}a_{B} \rangle = \langle a^{\dagger}_{S}a_{B}\rangle = \langle a^{\dagger}_{I}a^{}_{B}\rangle = \langle a^{\dagger}_{S}a^{\dagger}_{B}\rangle = \langle a^{\dagger}_{I}a^{\dagger}_{B}\rangle =0$, $\langle a_{I}a_{V}\rangle = \langle a^{\dagger}_{I}a^{}_{V}\rangle = \langle a^{\dagger}_{I}a^{}_{V}\rangle = \langle a^{\dagger}_{I}a^{\dagger}_{V}\rangle = \langle a^{\dagger}_{V}a^{}_{V}\rangle =0$, $ \langle a^{\dagger}_{S}a^{}_{S}\rangle = \langle a^{\dagger}_{I}a^{}_{I}\rangle = N_{S}$, $[a^{}_{S},a^{\dagger}_{S}]=[a^{}_{I},a^{\dagger}_{I}]= [a^{}_{V},a^{\dagger}_{V}]=1$, $\langle a_{S}a_{I} \rangle = \langle 0,0 \lvert (a^{}_{S} \cosh{r}+e^{i \varphi}\sinh(r)a^{\dagger}_{I})(a^{}_{I} \cosh{r}+e^{i \varphi}\sinh(r)a^{\dagger}_{S}) \lvert 0,0\rangle = \sinh(r)\cosh(r) = \sqrt{N_{S}(1+N_{S})}$, $\langle a^{\dagger}_{S}a^{\dagger}_{I} \rangle = \langle 0,0 \lvert (a^{\dagger}_{S} \cosh{r}+e^{-i \varphi}\sinh(r)a^{}_{I})(a^{\dagger}_{I} \cosh{r}+e^{-i \varphi}\sinh(r)a^{}_{S}) \lvert 0,0\rangle = \sinh(r)\cosh(r) = \sqrt{N_{S}(1+N_{S})}$, $\sinh^{2}(r)=N_{S}$, and we have set $\varphi=0$.

Then a similar calculation of the momentum translated output yields
\begin{align}
     \langle [Y^{(\text{out})}_{\mathrm M}]^{2} \rangle = \langle [Y^{2}_{R}-2GY_{R }Y_{\mathrm M}+G^{2}Y^{2}_{\mathrm M}] \rangle, 
     \label{eq:CNOTOUT2}
\end{align}
where,
\begin{align}
\langle Y^{2}_{R} \rangle&= \frac{-1}{4} \big \langle (\sqrt{\eta} a_{S}+\sqrt{1-\eta}a_{B}-\sqrt{\eta}a^{\dagger}_{S}-\sqrt{1-\eta}a^{\dagger}_{B}) \nonumber \\
    &\hspace*{1.5em}(\sqrt{\eta} a_{S}+\sqrt{1-\eta}a_{B}-\sqrt{\eta}a^{\dagger}_{S}-\sqrt{1-\eta}a^{\dagger}_{B}) \big \rangle \nonumber\\
&=\frac{1}{4} \big(  \eta [1 +2\langle a^{\dagger}_{S}a_{S}\rangle] +(1-\eta)[1 +2 \langle a^{\dagger}_{B}a_{B} \rangle]\big) \nonumber \\
&=\frac{\big(  \eta(1+ 2N_{S}) +(1-\eta)(1+2N_{B})\big)}{4} , \nonumber\\
\langle Y_{R}Y_{\mathrm M} \rangle &= \nonumber \\
&\frac{-1}{4} \big \langle (\sqrt{T} a_{I}+\sqrt{1-T}a_{V}-\sqrt{T}a^{\dagger}_{I}-\sqrt{1-T}a^{\dagger}_{V}) \nonumber \\ 
 &\hspace*{1.5em}(\sqrt{\eta} a_{S}+\sqrt{1-\eta}a_{B} -\sqrt{\eta}a^{\dagger}_{S}-\sqrt{1-\eta}a^{\dagger}_{B}) \big \rangle \nonumber \\
 &=\frac{-\sqrt{\eta T N_{S}(1+N_{S})}}{2}   
\end{align}
\begin{align}
\langle Y^{2}_{\mathrm M} \rangle &= 
    \frac{-1}{4}\big \langle (\sqrt{T} a_{I}+\sqrt{1-T}a_{V}-\sqrt{T}a^{\dagger}_{I} -\sqrt{1-T}a^{\dagger}_{V}) \nonumber \\ 
&\hspace*{1.5em}(\sqrt{T} a_{I}+\sqrt{1-T}a_{V}-\sqrt{T}a^{\dagger}_{I}-\sqrt{1-T}a^{\dagger}_{V}) \big \rangle \nonumber \\
&=\frac{(2TN_{S}+1)}{4}
 \label{eq:derv2}
\end{align}
As for the unaffected modes they are equal to,  $\langle [X^{(\text{out})}_{R}]^{2} \rangle = \langle X^{2}_{R} \rangle$, $\langle [Y^{(\text{out})}_{\mathrm M}]^{2} \rangle = \langle Y^{2}_{\mathrm M}\rangle$. It is clear now from Eqs. (\ref{eq:derv1}) and (\ref{eq:derv2}) that the receiver's translated modes $ \langle [X^{(\text{out})}_{\mathrm M}]^{2} \rangle$, $\langle [Y^{(\text{out})}_{R}]^{2} \rangle$ indeed carry the total signal-idler cross correlation signature $\langle X_{\mathrm M}X_{R} \rangle$, $\langle Y_{R} Y_{\mathrm M} \rangle$, nonetheless accompanied by unwanted noise. The receiver's output when $H=1$, is the sum of the signal powers of all the receiver's output quadratures 
\begin{align}
    \mathcal{I}_{1} &= \langle X^{2}_{\mathrm M} \rangle + \langle Y^{2}_{\mathrm M}\rangle+2G[\langle X_{\mathrm M }X_{R}\rangle-\langle Y_{R}Y_{\mathrm M} \rangle] \nonumber \\ &+ G^{2} [\langle X^{2}_{R}\rangle+\langle Y^{2}_{\mathrm M}\rangle] + \langle X^{2}_{R}\rangle +\langle Y^{2}_{R}\rangle +  \langle X^{2}_{V} \rangle+\langle Y^{2}_{V} \rangle
    \label{eq:IOUT1}
\end{align}
where the vacuum contribution stems from the  noise penalty on all measurements.

When the null hypothesis is true, $H=0$, the target return is replaced with a bath mode and the four receiver's outputs become 
\begin{align}
    X^{\text{(out)}}_{R} &= X_{B} \nonumber \\ 
    Y^{\text{(out)}}_{R} &= Y_{B}-GY_{\mathrm M} \nonumber \\
    X^{\text{(out)}}_{\mathrm M} &= X_{\mathrm M} +GX_{B}\nonumber \\
    Y^{\text{(out)}}_{\mathrm M} &= Y_{\mathrm M}
    \label{eq:Bath}
\end{align}
Then, 
\begin{align}
    \langle [X^{\text{(out)}}_{\mathrm M}]^{2} \rangle &= \langle X^{2}_{\mathrm M} \rangle +2G \langle X_{\mathrm M}X_{B} \rangle +G^{2} \langle X^{2}_{B} \rangle, \nonumber \\
   \langle [X^{(\text{out})}_{R}]^{2} \rangle &=\langle X^{2}_{B} \rangle = \frac{1}{2} \langle (a_{B}+a_{B}^{\dagger})(a_{B}+a_{B}^{\dagger})\rangle =\frac{(1+2N_{B})}{4} , \nonumber \\ 
\langle X^{2}_{\mathrm M} \rangle &=\frac{(2TN_{S}+1)}{4}, \nonumber \\
\langle [Y^{(\text{out})}_{R}]^{2} \rangle &= \langle Y^{2}_{B} \rangle -2G\langle Y_{B}Y_{\mathrm M} \rangle + G^{2}\langle Y^{2}_{\mathrm M} \rangle, \nonumber \\
  \langle Y^{2}_{B } \rangle &=\frac{-1}{4}\langle (a_{B}-a^{\dagger}_{B})(a_{B}-a^{\dagger}_{B}) \rangle \nonumber \\ 
 &=\frac{(1+2N_{B})}{4} ,\nonumber \\
\langle [Y^{\text{(out)}}_{\mathrm M}]^{2} \rangle &= \langle Y^{2}_{\mathrm M} \rangle  =\frac{(2TN_{S}+1)}{4}
\label{eq:Bath2}
\end{align}
where $\langle X_{\mathrm M}X_{B} \rangle = \langle Y_{B}Y_{\mathrm M} \rangle =0$, since the bath mode is not correlated with the stored idler.

Correspondingly, it can be seen that the receiver's output when the null hypothesis is true becomes 
\begin{align}
     \mathcal{I}_{0} &= \langle X^{2}_{\mathrm M} \rangle + \langle Y^{2}_{\mathrm M}\rangle+ G^{2} [\langle X^{2}_{B}\rangle+\langle Y^{2}_{\mathrm M}\rangle] + \langle X^{2}_{B}\rangle \nonumber \\ &+\langle Y^{2}_{B}\rangle +  \langle X^{2}_{V} \rangle+\langle Y^{2}_{V} \rangle
    \label{eq:IOUT2}
\end{align}
Since in the low brightness regime the following approximation is valid, $\langle X^{2}_{R} \rangle = \langle Y^{2}_{R} \rangle = \langle X^{2}_{B} \rangle = \langle Y^{2}_{B} \rangle$, it can be concluded that the effective signal power of the CNOT receiver is 
\begin{align}
    \mathcal{I}_{1}-\mathcal{I}_{0} \approx 2G\sqrt{\eta T N_{S}(N_{S}+1)}
    \label{eq:CNOTSIGPOWER}
\end{align}
In summary we have demonstrated the details of the process of extracting the signal-idler cross correlation in Eq. (\ref{eq:CNOTSIGPOWER}). This wil be the relevant quantity when we start discussing the receiver's error exponent. We have further shown that the receiver's output signal power is enhanced by the receiver's gain. Thus indeed the CNOT receiver can in principle offer a better performance than the other QI protocols. In order to quantify practically the amount of gain that can be controlled to enhance the detection process, we have to consider the effect of background noise on the device operation. This will be the task of the next section. 
\subsection{Background noise of CNOT receiver}
As shown in appendix \ref{sec:Analysis}, a double port homodyne measurement is capable of extracting the input field power, which is displayed on a spectrum analyzer's screen. In order to calculate the error exponent of our receiver, we need to calculate its noise power. This corresponds to calculating the PSD of a bath mode. Consider the case where the null hypothesis is true, that is, a returned mode is replaced with a bath one. Under the AWGN channel model, the bath mode enters the receiver as a white Gaussian random process, whereas the stored idler is an a thermal state with average photon $N_{S}$ after tracing out its signal twin. In such case the receiver output is Eq. (\ref{eq:Bath}).
In the low brightness regime we can approximately neglect the power of the memory mode, and that corresponding to the vacuum noise penalty, thus we calculate the bath quadrature noise power according to Eqs. 
 (\ref{eq:homd1}-\ref{eq:homd2}) as
\begin{align}
    \big \langle  X_{B}^{2} \big \rangle  &\approx \frac{\langle (a_{B}+a^{\dagger}_{B})(a_{B}+a^{\dagger}_{B})\rangle}{4} \nonumber \\
    & \approx \frac{1+2N_{B}}{4} \approx \frac{N_{B}}{2 }
\end{align}
Similarly, the noise power of the momentum quadrature is calculated as 
\begin{align}
    \big \langle  Y_{B}^{2} \big \rangle &\approx \frac{-\langle (a_{B}-a^{\dagger}_{B})(a_{B}-a^{\dagger}_{B})\rangle}{4} \nonumber \\
    & \approx \frac{1+2N_{B}}{4} \approx \frac{N_{B}}{2 }
\end{align}
where in the above equations we have used the bath properties $\langle a_{B} \rangle = \langle a^{\dagger}_{B} \rangle = 0$, and we assumed that powers are measured in a narrow bandwidth.

Thus the overall noise power becomes 
\begin{align}
    \mathrm{P}_{\mathrm N}(\mathcal{I}_{0}) &= \big \langle  [{X^{\text{(out)}}_{R}}]^{2} \big \rangle + \big \langle [{Y^{(\text{out})}_{R}}]^{2} \big \rangle+\langle  [{X^{(\text{out})}_{\mathrm M}}]^{2} \big \rangle \nonumber \\ 
    &+ \langle  [{Y^{(\text{out})}_{\mathrm M}}]^{2} \big \rangle  \nonumber \\
    &= G^{2}\langle  X^{2}_{B} \rangle +  \langle  X^{2}_{\mathrm M} \rangle + \langle  X^{2}_{B} \rangle+ \langle  Y^{2}_{\mathrm M} \rangle \nonumber \\
    &+G^{2}\langle  Y^{2}_{\mathrm M} \rangle+\langle  Y^{2}_{B} \rangle
    \end{align}
    Thus, 
    \begin{align}
   \mathrm{P}_{\mathrm N}(\mathcal{I}_{0}) & \approx N_{B} + \frac{G^{2}N_{B}}{2} = N_{B}(1+\frac{G^{2}}{2})
\end{align}
where we recall that $\langle X_{B}\rangle = \langle Y_{B}\rangle =\langle X_{\mathrm M}\rangle = \langle Y_{\mathrm M}\rangle=0$.\\

 Since the background noise is identical for both transmission hypotheses, we will assume equal hypotheses noise power, $\mathrm{P}_{\mathrm N}(\mathcal{I}_{0})=\mathrm{P}_{\mathrm N}(\mathcal{I}_{1})=\mathrm{P}_{\mathrm N}$. This is a reasonable approximation in communication systems, when a thermal bath is the dominant noise source \cite{desurvire1995erbium}. Hence the device background noise power is 
\begin{align}
    \mathrm{P}_{\mathrm {N}} & \approx N_{B}(1+\frac{G^{2}}{2})
    \label{eq:ADDNOISE}
\end{align}
The previous expression is used to calculate the receiver's error exponent as shown in the next section.
\section{Performance analysis}
\begin{figure*}
    \centering
   \subfloat[]{\includegraphics[scale=0.33]{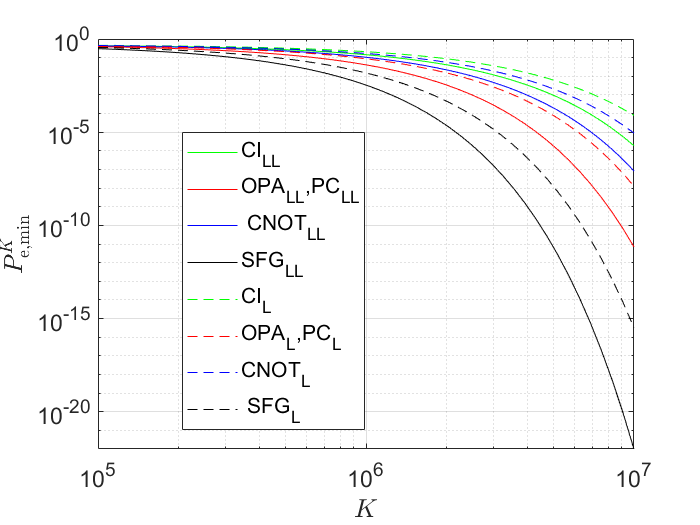}}
   \subfloat[]{\includegraphics[scale=0.33]{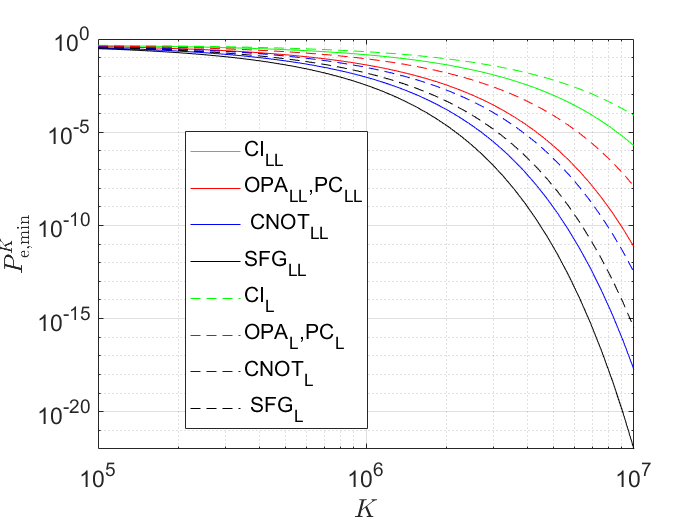}}
   \subfloat[]{\includegraphics[scale=0.33]{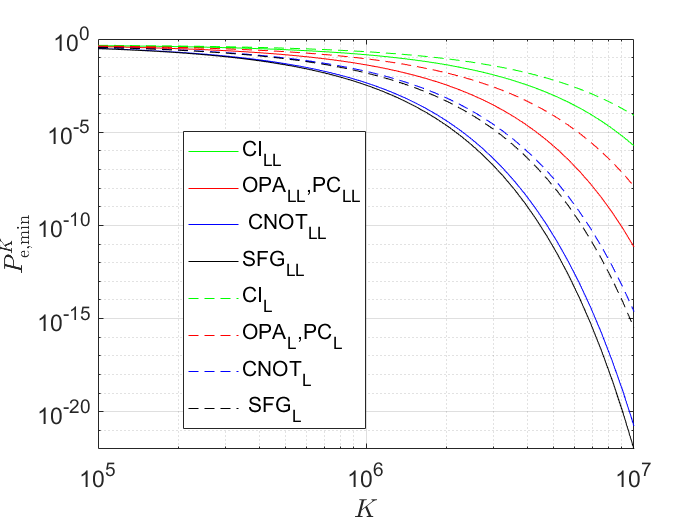}}
    \caption{A plot of the minimum bit error probability of all QI receivers against the total number of probe pairs $K$. We have also included the optimal CI receiver  in order to demarcate the domain of the quantum advantage. The subscript "LL" denotes the ideal lossless operation, whereas "L" means that the transmissivity of the lossy idler storage is non ideal. In this plot it was chosen to be $T=0.7$, since according to \cite{flurin2015superconducting}, a realistic microwave quantum memory has an efficiency of $80 \%$. In Fig. (\ref{fig:Error}a). the CNOT is operating in the unity gain regime with a beamsplitter parameter $g \approx 0.38$. As can be seen, the SFG receiver has the best performance among all receivers in both the lossless and lossy operations. The unity gain CNOT only outperforms the optimal CI receiver, while being outperformed by all QI receivers in both the LL and L operations. In order to achieve an enhanced performance over OPA and PC in both the LL and L operations, we observed that a gain of $G=1.5$ is enough for the task. In Fig. (\ref{fig:Error}b) the CNOT receiver operates with a larger gain value, specifically, $G=3$. The corresponding beamsplitter parameter in this case is $g \approx 0.09$. In this regime of operation, the L CNOT receiver (dotted blue) outperforms the LL PC and LL OPA receivers (solid red ). On the other hand, it can be seen that LL CNOT slightly outperforms L SFG while being outperformed by the LL SFG. Thus, SFG still has the best performance as before. We also note that the optimum CI has the least performance both in the LL operation (solid green) and L operation (dotted green). In Fig. (\ref{fig:Error}c),  the CNOT receiver operates with a gain equal to $G=6$, and a corresponding beamsplitter parameter of $g \approx 0.03$. It is clear that the CNOT performance is only comparable to SFG in both LL and L cases, while closing the performance gap, still is outperformed by SFG. Nonetheless, the L CNOT managed to outperform both the LL OPA and LL PC. For completeness, we observe that similar as before, the optimum CI has the least performance in both operations LL and L.  }
    \label{fig:Error}
\end{figure*}
The objective of this section is to demonstrate the operational regime where our device can outperform other QI receivers. Theoretically we have shown in the last section that our receiver's gain can indeed strengthen the signal-idler cross correlations, which in principle should translate into a better device SNR. However, practically the device internal interactions add extra noise to that in the channel. Thus it is imperative to study the effect of the overall noises in the system on the performance of our device. We will use the practical setup described in appendix \ref{Practical} as our model of device noise, this will help us understand exactly how the receiver's gain can be manipulated to achieve the desired enhancement. This section is organized as follows: we begin with a simplified background on the basics of error probability in transmission problems tailored to the QI scenario. Then we derive the error probability formula of our device. Finally we plot our device's error bounds in different device settings and compare them to the other QI protocols in order to highlight our areas of improvement.

A good performance metric of a QI receiver is its ability to circumvent high error rates when discriminating between the two possible transmission hypotheses. This binary decision situation is identical to on-off communications systems \cite{gallager2013stochastic}, such that when $H=1$ is true, the 1-bit signal ($j_{1} $) is sent, whereas when $H=0$ is true the 0-bit ($j_{0}$) signal is sent. Further, we suppose that the receiver's environment, that is, the channel's noise, is \textit{additive white Gaussian noise} (AWGN). Empirically this is a reasonable assumption in most practical cases, since the electrons random motion inside the receiver's front end conductors is modelled as a \textit{stationary Gaussian random process}.  The receiver's total \textit{bit error rate} (BER) \cite{agrawal2012fiber} is then defined as $P_{e} = p(1)p(0 \lvert 1)+p(0)p(1 \lvert 0)$, where $p(1)$ (prior) is the probability that the target is there, $p(0 \lvert 1)$ (conditional) is the probability of a \textit{miss}, i.e., deciding that the target is not there while in reality it is there, $p(0)$ is the probability that the target is not there, $p(1 \lvert 0)$ is the probability of a \textit{false alarm}, i.e.,  deciding that target is there while in reality it is not there. The conditionals are calculated with respect to a decision threshold $j_{d}$ as $p(0 \lvert 1) = (1/\sqrt{2 \pi \sigma^{2}_{1}})\underset{-\infty}{\overset{j_{d}}{\int}} \exp{\big(-(j-j_{1})/2 \sigma^{2}_{1}\big)} dj = (1/2) \text{erfc}\big((j_{1}-j_{d})/\sqrt{2} \sigma_{1}\big)$,\\ $p(1 \lvert 0) =  (1/\sqrt{2 \pi \sigma^{2}_{0}})\underset{j_{d}}{\overset{\infty}{\int}}\exp{\big(-(j-j_{0})/2 \sigma^{2}_{0}\big)} dj = (1/2 )\text{erfc}\big((j_{d}-j_{0})/\sqrt{2} \sigma_{0}\big)$, where the means of the $0 \& 1 $ bit signals are $j_{0}$, $j_{1}$ respectively, whereas $\sigma^{2}_{0}, \sigma^{2}_{1}$ are the \textit{filtered} power spectral density (PSD) of the zero mean white Gaussian noise process comprising the receiver's environment when $H=0$, $H=1$ respectively, and $\text{erfc}(x) =(2/\pi) \underset{x}{\overset{\infty}{\int}}e^{-y^{2}}dy $ is the complementary error function. Assuming equal priors, $p(0)=p(1)=1/2$, BER reads $P_{e} = 1/4 \big[ \text{erfc}\big((j_{1}-j_{d})/\sqrt{2} \sigma_{1}\big)+\text{erfc}\big((j_{d}-j_{0})/\sqrt{2} \sigma_{0}\big)\big]$. We note that a filtered zero mean white Gaussian random process has a variance equals to its PSD, this complies with empirical observations. The BER minimum occurs when $j_{d}$ is chosen such that, $(j_{d}-j_{0})^{2}/2\sigma^{2}_{0} = (j_{1}-j_{d})^{2}/2\sigma^{2}_{1} + \text{ln} \big( \sigma_{1}/ \sigma_{0} \big)$. Under the assumption that the noise PSD is equal for both hypotheses, $\sigma_{0}=\sigma_{1}=\sigma$, we arrive at an expression for the decision threshold as $(j_{d}-j_{0})/\sigma = (j_{1}-j_{d})/\sigma = R_{Q}$, $j_{d} =(j_{0}+j_{1})/2$, where $R_{Q}$ is denoted by the error exponent. An expression for the error exponent can be written as $R_{Q}= (j_{1}-j_{0})/2\sigma$. A more general form of the error exponent is adopted when the noise PSD is different for the two transmission hypotheses, $R_{Q} =(j_{1}-j_{0})/(\sigma_{0}+\sigma_{1})$. Moreover, the minimum BER as a function of the error exponent can be provided as, $P_{\text{e, min}} = (1/2) \text{erfc}\big(\frac{R_{Q}}{\sqrt{2}}\big)$. By exploiting a series expansion of the \text{error} function, the minimum BER can be written as $P_{\text{e, min}} = \frac{1}{R_{Q}\sqrt{2\pi}} \exp{\big(\frac{-R^{2}_{Q}}{2}\big)}+\frac{R_{Q}}{2\sqrt{2\pi}}\big[ \frac{\sqrt{2 \pi}}{R_{Q}}-2- \underset{n=1}{\overset{\infty}{\sum}} \big( \frac{-R^{2}_{Q}}{2} \big)^{n} \frac{1}{(n+1)!(2n+1)!}\big]$, such that it can be approximated as, $P_{\text{e, min}} \approx \frac{1}{R_{Q}\sqrt{2\pi}} \exp{\big(\frac{-R^{2}_{Q}}{2}\big)} $. A practical upper bound on the minimum error probability neglects the denominator of this expression, as we shortly see. After this brief motivation, the error exponent of the CNOT receiver can be defined as
 

\begin{align}
& R_{Q_{\text{CNOT}}} = \frac{R_{Q}^{2}} {2}= \frac{1}{2}\Bigg[\frac{\mathcal{I}_{1}-\mathcal{I}_{0}}{  \sqrt{\mathrm{P}_{\mathrm N}(\mathcal{I}_{1})}+\sqrt{\mathrm{P}_{\mathrm N}(\mathcal{I}_{0})}}\Bigg]^{2},
    \label{eq:errorCNOT}
\end{align}
where $\mathcal{I}_{1}$ is the average receiver's output when $H=1$,  given by the expression in Eq. (\ref{eq:IOUT1}), whereas $\mathcal{I}_{0}$ is given by \ref{eq:IOUT2}, which is the receiver's output when the null hypothesis is true. Their associated noise powers are defined as $\mathrm{P}_{\mathrm N}(\mathcal{I}_{1})$, $\mathrm{P}_{\mathrm N}(\mathcal{I}_{0})$ respectively. Similarly, for equal noise powers we define the $\text{SNR}_{\text{CNOT}}$ as $4 R_{Q_{\text{CNOT}}}$.

Further, it has been shown that for $K$ probe signals, the minimum bit error probability  is upper bounded by the  classical Bhattacharyya bound \cite{guha2009gaussian}  
\begin{align}
   P^{K}_{e, \text{min}} \leq \frac{1}{2}\exp{[-KR_{Q}]},
   \label{eq:boundOPA}
\end{align} 
where $K$ is the total number of signal-idler pairs generated at the transmitter.

We are now ready to compare between the error probability upper bounds of different QI receivers. Following \cite{guha2009gaussian}, and \cite{shapiro2020quantum}, the error exponents of the OPA, PC and SFG are
\begin{align}
    R_{Q_{\text{OPA, PC}}} = \eta T N_{S}/ 2N_{B}
    \label{eq:SNROPA}
\end{align}
\begin{align}
    R_{Q_{\text{SFG}}} = \eta T N_{S}/ N_{B}
    \label{eq:SNRSFG}
\end{align}
respectively.

For the CNOT receiver, we assume that both hypotheses have equal noise power based on the analysis presented in appendix \ref{sec:Analysis}. Thus its error exponent expression according to Eq. (\ref{eq:errorCNOT}) becomes, 
\begin{align}
    R_{Q_{\text{CNOT}}} = \eta G^{2}T N_{S}/ 2 \mathrm{P}_{\mathrm N}
    \label{eq:SNRCNOT}
\end{align}
where $\mathrm{P}_{\mathrm N}(\mathcal{I}_{0})=\mathrm{P}_{\mathrm N}(\mathcal{I}_{1})=\mathrm{P}_{\mathrm N}(\mathcal{I})$ is defined by Eq. (\ref{eq:ADDNOISE}).
We further assumed for all receivers that $\eta = 0.01$, $T=0.7$, $N_{S} =0.01$, and $N_{B} =20$.

 In Fig. (\ref{fig:Error}) we have plotted the minimum bit error probability against the total number of probe pairs. We have also included the the optimum CI receiver to demonstrate the quantum advantage in the low SNR setting. Let us consider first the trivial case of zero gain operation. In the event of zero interaction gain, $G=0$, the CNOT receiver homodynes both the return and the stored idler individually. Consequently, the $3\text{dB}$ noise penalty due to the simultaneous measurement of the two non commuting quadratures eradicates any quantum advantage. We now focus on non zero gain operation of our receiver. We considered three different gain values for our CNOT receiver, namely, unity gain,  $G=3$, and $G=6$.  By substituting with $G=1$ in Eq. (\ref{eq:ADDNOISE}), it can be seen that the total number of added noise photons in this case is $64$. As can be seen from Fig. (\ref{fig:Error}a), the performance of the CNOT receiver was only able to outperform the optimum CI in both the LL operation, that is, $T=1$ and the L operation, that is, $T=0.7$. Nonetheless, it was outperformed by all QI receivers in both cases respectively. We note that the SFG receiver has the best performance among all receivers in both operations for this case. Further, the unity gain case is interesting in itself, since it represents the domain of operation where the device operates typically as a qubit CNOT gate. Thus we conclude that any other realization of the CNOT operation based on a different platform would replicate the same performance.

In Fig. (\ref{fig:Error}b) the CNOT receiver operates with a gain above unity, i.e., $G=3$. The total number of added noise photons is $\approx 224$. In this domain of operation it can be seen that the CNOT outperforms both the OPA and PC in the LL and L cases respectively. However, it is still being outperformed by the SFG receiver. 

In Fig. (\ref{fig:Error}c) the CNOT receiver operates with a gain equal to $G=6$. The total number of added noise photons in this case is $\approx 764$. It can be seen from the plot, that in this case the CNOT receiver is only comparable to the SFG, although, still being outperformed by it in the LL and L cases respectively. We further observe that the CNOT outperformed both LL OPA and LL PC even when operating in its lossy operation. Thus, we can conclude from the plots in Fig. (\ref{fig:Error}), that by increasing the CNOT gain its performance approaches that of the SFG. Further, by analyzing Eq. (\ref{eq:ADDNOISE}) in the limit of large gain, $G>>1$, and assuming negligible internal noise, that is, strong squeezing and negligible homodyne detection inefficiencies, the variance of the receiver's output becomes $\text{Var}(\mathcal{I}) \approx N_{B}G^{2}/2$, and consequently the error exponent would be $R_{Q_{\text{CNOT}}} = \eta T N_{S}/N_{B}$, therefore coinciding with that of SFG. 
\section{Conclusion}
In this paper we have considered a new QI receiver design for microwave applications. Due to the technological difficulty of realizing single photon counters, the proposed device relies completely on homodyne measurements and square law detectors. The receiver is built upon an offline controlled gain CV CNOT gate in order to extract the signal-idler cross correlations. 

We have investigated different gain operational values of our CNOT receiver. In the unity gain scenario, we have shown that the CNOT offers no performance advantage over any of the QI receivers, while only managing to edge past the optimum CI receiver. We expect similar performance from any other realization of a  unity gain CNOT gate. On the other hand when operating with above unity gain, we showed that our device approached the best QI receiver gradually as the gain increases. Ideally when squeezing and vacuum noises are suppressed, a high gain operating point matches the SFG receiver. We further noticed that with squeezing noise, an above unity gain CNOT can still offer a decent performance, comparable to SFG, especially in the radar domain, where the maximum number of utilised probe pairs is $\approx 10^{5}-10^{6}$ \cite{lanzagorta2011quantum}. This is visible in the error probability curves in Figs. (\ref{fig:Error}b), and (\ref{fig:Error}c). 

Two final remarks on the engineering challenges of implementing the protocol in the microwave domain. Tailoring a desired high gain operational point requires a small and controllable beamsplitter coefficient $g$ by virtue of the relation $G=(1-g)/\sqrt{g}$. Recently significant progress has been made towards engineering devices capable of achieving this level of controlled transformations \cite{hoffmann2010superconducting, lu2023high}. Further, we have also observed that a high gain operating point is usually accompanied by excess noise photons, this may result in an elongated dead time of our receiver; however, recent techniques in cQED can mediate excess noise by utilising circuit refrigeration procedures \cite{tan2017quantum}. These are all clear signs that the proposed model can be practically implemented by the existing quantum microwave technologies.
\begin{appendices}
\section{}
\section*{Analysis of CNOT receiver's homodyne measurement}\label{sec:Analysis}
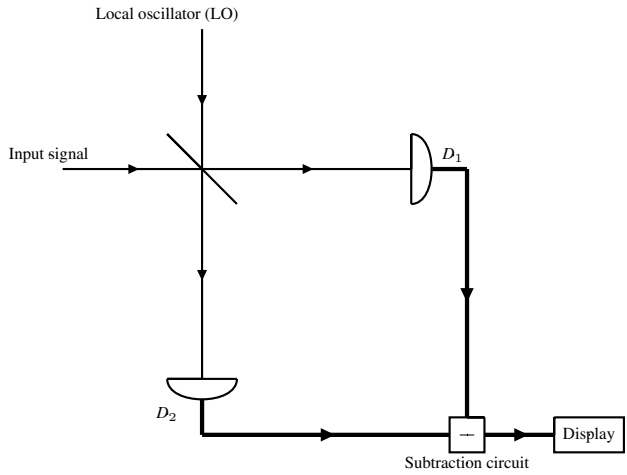
\begin{figure}
    \centering
    \begin{tikzpicture}[thick,scale=0.92, every node/.style={scale=0.9}]
        \draw[thick] (0,0)--(2,0) node [currarrow, pos=0.5, scale=0.7] {};
        \draw[thick] (2,0)--(1.5,0.5)--(2.5,-0.5);
        \draw[thick] (2,0)--(5,0) node [currarrow, pos=0.5, scale=0.7] {};
        \draw[thick] (5,0)--(5,0.5)--(5,-0.5);
        \draw[thick](5,0.5) to [bend left = 90] (5,-0.5);
        \draw[thick] (2,2)--(2,0) node [currarrow, pos=0.5, sloped, scale=0.7] {};
        \draw[thick] (2,0)--(2,-3) node  [currarrow, pos=0.5, sloped, scale=0.7] {};
        \draw[thick](2,-3)--(1.5,-3)--(2.5,-3);
        \draw[thick](1.5,-3) to [bend right = 90] (2.5,-3);
        \draw[ultra thick] (2,-3.3 )--(2,-3.8);
        \draw[ultra thick] (5.3,0 )--(5.8,0);
        \draw[ultra thick] (5.8,0)--(5.8,-3.55)  node [currarrow, pos=0.5, sloped, scale=0.7]{};
        \draw[thick] (5.8,-3.55)--(6.05,-3.55)--(5.55,-3.55)--(5.55,-4.05)--(6.05,-4.05)--(6.05,-3.55);
        \draw[ultra thick]  (2,-3.8 )--(5.55,-3.8)  node [currarrow, pos=0.5, sloped, scale=0.7]{};
        \draw[] (5.8,-3.8)--(5.8,-3.8) node []{$-$};
        \draw[ultra thick]   (6.05,-3.8)--(7.05,-3.8)  node [currarrow, pos=0.5, sloped, scale=0.7] {};
        \draw[thick] (7.05,-3.8)--(7.05, -3.55)--(7.05, -4.05)--(8.05,-4.05)--(8.05,-3.55)--(7.05,-3.55);
        \draw[] (7.55, -3.8)--(7.55,-3.8) node []{\scriptsize $\text{Display}$};
        \draw [] (-0.2,0.2)--(-0.2,0.2) node []{\scriptsize \text{Input signal}};
        \draw [] (1.5,2.2)--(1.5,2.2) node [] {\scriptsize \text{Local oscillator (LO)}};   
        \draw[](5.6,0.2)--(5.6,0.2) node [] {\scriptsize $D_{1}$};
        \draw[](1.5, -3.5)--(1.5,-3.5) node [] {\scriptsize $D_{2}$};
        \draw [] (5.8, -4.2)--(5.8,-4.2) node []{\scriptsize \text{Subtraction circuit}};
    \end{tikzpicture}
    \caption{A schematic of the homodyne detection chain used by our CNOT receiver. The setup is composed of a balanced beamsplitter and two detetctors $D_{1}, D_{2}$. The desired signal is injected from the beamsplitter's first port, while a strong local oscillator (LO) field enters from the second. The detectors outputs are directed towards a subtraction circuit in order to display the final result. The double port homodyne circuit can be thought of as being embedded inside a spectrum analyzer device, such that its display shows the power of the measured input. }
    \label{fig:homCircuit}
\end{figure}
The CNOT receiver as described in the main text operates on the fields non-commuting quadrature operators. The noise penalty of measuring two non-commuting observables is 3dB. This can be seen when we consider splitting individually each of the returned mode and the stored idler, $a_{R}, a_{\mathrm M}$, on a balanced beamsplitter, where a vacuum mode enters  from its unused port, $\bar{a}_{R}= (1/ \sqrt{2})(a_{R}-ia_{v})$, $\bar{a}_{v} =(1/\sqrt{2})(a_{v} +ia_{R})$, such that $\bar{X}_{R} = (1/ \sqrt{2})(\bar{a}_{R}+\bar{a}^{\dagger}_{R}) = (1/4)(a_{R}-ia_{v}+a^{\dagger}_{R}+ia^{\dagger}_{v})$, and $\bar{Y}_{v} = (1/ i\sqrt{2})(\Bar{a}_{v}-\bar{a}^{\dagger}_{v})=(1/i4)(a_{v}+ia_{R}-a^{\dagger}_{v}+ia^{\dagger}_{R})$, then $[\Bar{X}_{R}, \Bar{Y}_{v}] = (1/i16)\big(i[a^{}_{R},a^{\dagger}_{R}]+i[a^{\dagger}_{R},a^{}_{R}]+i[a^{}_{v},a^{\dagger}_{v}]+i[a^{\dagger}_{v},a^{}_{v}]\big)=0$, and hence these two observables can be measured simultaneously \cite{haus2000electromagnetic}. However, as can be seen, the noise penalty due to splitting the mode first on a balanced beamsplitter is present as an attenuation factor of $1/2$, where only half of the original intensity is contained in $\bar{X}_{R}$, $\bar{Y}_{V}$. One might wonder now that the cross correlation output of our CNOT receiver would suffer a similar fate. This would have been the case if our receiver measures the return and the stored modes individually without mixing them first. However the interaction between the two modes as described by the gate transformations in Eq. (\ref{eq:CNOTRXOUT}) and the analysis that followed, showed that the cross correlation signature was preserved. 

We now focus on outlining the details of our receiver's homodyne chain. 
Following \cite{schumaker1984noise, ukai2014multi}, Fig. (\ref{fig:homCircuit}) presents a schematic of the detection circuit used by our receiver to output the measured values of the observable quantities in Eqs. (\ref{eq:CNOTOUT1}) and (\ref{eq:CNOTOUT2}) (see also Fig. (\ref{fig:Unitary})). As can be seen, the input field is mixed on a balanced beamsplitter with a local oscillator field followed by two detectors, $D_{1}, D_{2}$, a subtraction circuit that calculates the difference between the generated photo-currents, and a a spectrum analyzer display that shows the measured field's variance (power). Without loss of generality, let's consider an arbitrary returned mode $a_{R}$, not necessarily in a TMSV with an idler. We further assume a noiseless transmission of this return. On the other hand, at the receiver, our local oscillator is tuned to extract the unaffected quadrature $Y_{R}$. The output of the detection chain is defined as follows, \begin{align}
    a^{\text{(out)}}_{R} &= \frac{1}{\sqrt{2}}(a^{\text{(in)}}_{R}+i d^{\text{(in)}}_{}) \nonumber \\
     d^{\text{(out)}}_{} &= \frac{1}{\sqrt{2}}(d^{\text{(in)}}_{}-i a^{\text{(in)}}_{R})
\end{align} 
where $d^{\text{(in)}}$ is the local oscillator mode.

Then, the output of the subtraction circuit is, 
\begin{align}
    \mathrm I &=  {a^{\text{(out)}}_{R}}^{\dagger} a^{\text{(out)}}_{R}- {  d^{\text{(out)}}_{}}^{\dagger}  d^{\text{(out)}}_{}, \nonumber \\
    N^{\text{(out)}}_{R}&=  \frac{1}{2}(N^{\text{(in)}}_{R}+N^{\text{(in)}}_{d}+i{d^{\text{(in)}}}^{\dagger}a^{\text{(in)}}_{R}-i{d^{\text{(in)}}}{a^{\text{(in)}}_{R}}^{\dagger}), \nonumber \\
    N^{\text{(out)}}_{d}&= \frac{1}{2}(N^{\text{(in)}}_{R}+N^{\text{(in)}}_{d}-i{d^{\text{(in)}}}^{\dagger}a^{\text{(in)}}_{R}+i{d^{\text{(in)}}}{a^{\text{(in)}}_{R}}^{\dagger}) \nonumber \\
    \mathrm I &= i ({d^{\text{(in)}}}^{\dagger}a^{\text{(in)}}_{R}-{d^{\text{(in)}}}{a^{\text{(in)}}_{R}}^{\dagger})
\end{align}
where $N^{\text{(in)}}_{R} = {a^{\text{(in)}}_{R}}^{\dagger} a^{\text{(in)}}_{R}$, and $N^{\text{(in)}}_{d} = {d^{\text{(in)}}}^{\dagger}d^{\text{(in)}}$.

By assuming that the local oscillator mode is a complex number, $d^{\text{(in)}} \rightarrow \tilde{D}=\lvert \alpha_{L} \lvert e^{i \phi_{L}}$, we can extract the field's $Y$ quadrature by setting the LO phase to $\pi$ and normalizing the output current, 
\begin{align}
    Y_{R}&=\frac{\mathrm I}{\lvert\alpha _{L} \lvert \sqrt{2}} = \frac{-i (a_{R}-a^{\dagger}_{R})}{\sqrt{2}} 
\end{align}
where $\lvert \alpha_{L} \lvert$ is the LO field strength, and $\phi_{L}$ is its phase. 

One of the powerful features of double port homodyning is that the subtraction circuit eliminated the noise associated with the LO field. This results in the homodyned output noise power being only dependent on the input's variance, as we shall see now.  In order to estimate the overall noise accompanied with the process of double port homodyning \cite{schumaker1984noise}, we split the returned mode into a signal carrying part plus fluctuations, $a_{R} = \langle a_{R} \rangle+\Delta a_{R}$, such that $\langle a_{R} \rangle = A_{R}, \langle \Delta a_{R} \rangle = 0$, where $A_{R} = A^{X}_{R}+iA^{Y}_{R}$, $\Delta a_{R} = \Delta a^{X}_{R}+i\Delta a^{Y}_{R}$, $A^{X}_{R}, A^{Y}_{R}$ are the $X$, and $Y$ quadrature amplitude values respectively, and $\Delta a^{X}_{R}, \Delta a^{Y}_{R}$ are their associated fluctuations. Thus 
\begin{align}
    \langle \mathrm I \rangle &= i\lvert \alpha_{L} \lvert  (\langle a_{R} \rangle^{*}+ \langle \Delta a^{\dagger}_{R} \rangle- \langle a_{R} \rangle-\langle \Delta a_{R} \rangle)  \nonumber \\
    &= i  \lvert \alpha_{L} \lvert(A^{*}_{R}-A^{}_{R}) = 2i \lvert \alpha \lvert \hspace*{0.3em}{\text{Im}} [A^{*}_{R}] \nonumber \\
    &= 2\lvert \alpha_{L} \lvert A^{Y}_{R}
    \label{eq:homd1}
    \end{align}
    \begin{align}
    \langle \Delta {\mathrm I}^{2} \rangle &= \langle {\mathrm I}^{2} \rangle-\langle \mathrm I \rangle^{2}, \nonumber \\
    \langle \mathrm I^{2} \rangle &=\big \langle \big [ i \lvert \alpha_{L}\lvert (A^{*}_{R} +\Delta a^{\dagger}_{R} -A_{R}-\Delta a_{R}) \big ]^{2} \big \rangle \nonumber\\
    &=  \big \langle \big [ 2i \lvert \alpha_{L}\lvert \big({\text{Im}} [A^{*}_{R}]+ {\text{Im}}[\Delta a^{\dagger}_{R}]\big) \big ]^{2} \big \rangle\nonumber \\
    &= 4 \lvert \alpha_{L} \lvert^{2} \big \langle \big ( A^{Y}_{R}+\Delta  a^{Y}_{R}\big)^{2} \big \rangle \nonumber \\
    &\approx 4 \lvert \alpha_{L} \lvert^{2} {A^{Y}_{R}}^{2}+ 4 \lvert \alpha_{L} \lvert^{2} \langle \Delta {a^{Y}_{R}}^{2} \rangle , \nonumber \\
    \langle \Delta \mathrm I^{2} \rangle & \approx 4 \lvert \alpha_{L} \lvert^{2} \langle \Delta {a^{Y}_{R}}^{2} \rangle  \nonumber \\
    \langle \Delta Y_{R}^{2} \rangle &\approx \frac{\langle \Delta \mathrm I^{2} \rangle}{ 4 \lvert \alpha_{L} \lvert^{2} } = \langle \Delta {a^{Y}_{R}}^{2} \rangle
    \label{eq:homd2}
\end{align}
The above expressions show that balanced double port homodyning can extract both the mean and second moment (power) of a returned mode. Consider now the double port homodyning of a target return that is a part of a TMSV generated at the transmitter. Since our protocol operates in the microwave domain, the  detectors that produce $N^{(\text{out)}}_{R}$, and $N^{(\text{out})}_{d}$ respectively are \text{square law detectors} \cite{haus2000electromagnetic}, such as \textit{bolometers} \cite{lee2020graphene, walsh2017graphene, kokkoniemi2020bolometer}, for instance. Unlike single photon counters, the detector's medium in the case of a square law detector responds to the incident signal power. On the other hand, in single photon counters it  responds to the incident photon intensity or flux. Thus the former is a scalar quantity, while the later is a vector one.  

As pointed out in the main text, the expected value of the quadratures of a squeezed vacuum field vanish, that is to say, the average of the current generated after the subtraction circuit is zero, $\langle \mathrm I \rangle =0$. However, the variance of a zero mean squeezed vacuum field is non-zero. Thus we seek a device that can display these variances. This can be achieved by a spectrum analyzer, since in the case of a TMSV, the field variance of the input coincide with the field's second moment, i.e., its power,  $\langle \Delta {\mathrm I}^{2} \rangle = \langle {\mathrm I}^{2} \rangle$ as shown in Eq. (\ref{eq:homd2}). Thus the spectral output of the spectrum analyzer is proportional to the input field power. In summary, the homodyne measurement chain deployed by our CNOT receiver is composed of two steps; first the balanced double port homodyning captures the variance of the input signal, while suppressing the LO noise. Hence the detection noise is forced to be shot limited. Then the spectrum analyzer displays the measured power. It is worth mentioning that modern spectrum analyzer devices have a built in double port homodyne circuit and displays the input power at the end of the measurement.

We consider a similar process to extract the rest of the gates outputs. For the sake of completeness we show this for the other unaffected quadrature, that is, the memory mode position quadrature. While the rest of the outputs are just a linear superposition of the return and memory modes and can be deduced similarly in a straight forward manner. Consider now performing a double port homodyne measurement on the memory mode to extract its position quadrature. Similarly as before, the mode transforms at the detection chain as 
\begin{align}
     a^{(\text{out})}_{\mathrm M} &= \frac{1}{\sqrt{2}}(a^{\text{(in)}}_{\mathrm M}+i d^{\text{(in)}}_{}) \nonumber \\
     d^{\text{(out)}}_{} &= \frac{1}{\sqrt{2}}(d^{\text{(in)}}_{}-i a^{\text{(in)}}_{\mathrm M})
\end{align}
Then, the output of the subtraction circuit is, 
\begin{align}
    \mathrm I &=  {a^{\text{(out)}}_{\mathrm M}}^{\dagger} a^{\text{(out)}}_{\mathrm M}- {  d^{\text{(out)}}_{}}^{\dagger}  d^{\text{(out)}}_{}, \nonumber \\
    N^{\text{(out)}}_{\mathrm M}&=  \frac{1}{2}(N^{\text{(in)}}_{\mathrm M}+N^{\text{(in)}}_{d}+i{d^{\text{(in)}}}^{\dagger}a^{\text{(in)}}_{\mathrm M}-i{d^{\text{(in)}}}{a^{\text{(in)}}_{\rm M}}^{\dagger}), \nonumber \\
    N^{\text{(out)}}_{d}&= \frac{1}{2}(N^{\text{(in)}}_{\mathrm M}+N^{\text{(in)}}_{d}-i{d^{\text{(in)}}}^{\dagger}a^{\text{(in)}}_{\mathrm M}+i{d^{\text{(in)}}}{a^{\text{(in)}}_{\mathrm M}}^{\dagger}) \nonumber \\
    \mathrm I &= i ({d^{\text{(in)}}}^{\dagger}a^{\mathrm(in)}_{\mathrm M}-{d^{\text{(in)}}}{a^{\text{(in)}}_{\mathrm M}}^{\dagger})
\end{align}
Thus we can extract the field's $X$ quadrature by setting the LO phase to $\pi /2$ and normalizing the output current, 
\begin{align}
    X_{\mathrm M}&=\frac{\mathrm I}{\lvert\alpha _{L} \lvert \sqrt{2}} = \frac{ (a_{\mathrm M}+a^{\dagger}_{\mathrm M})}{\sqrt{2}} 
\end{align}
The field's power can be extracted as described before.

\section{}
\section*{ Practical model of the CNOT receiver} \label{Practical}
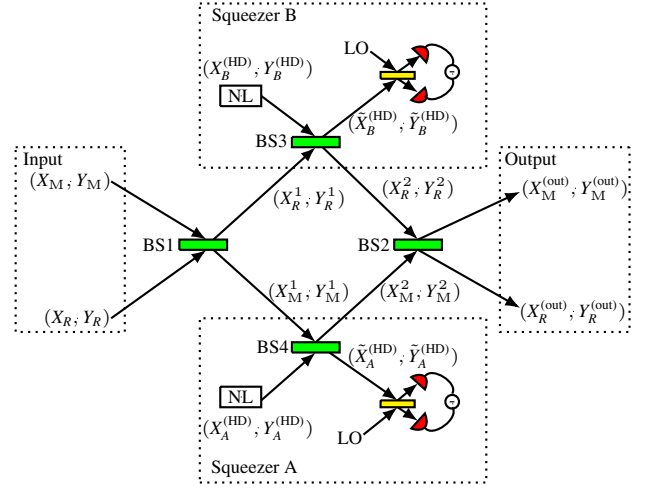
\begin{figure}
    \centering
    \begin{tikzpicture} [scale=0.9]
        \draw[thick, fill=green](0,0)--(0.7,0)--(0.7,0.15)--(0,0.15)--(0,0);
        \draw[thick, -{Latex[length=2mm]}] (-1,1)--(0.4,0.15);
        \draw[thick, -{Latex[length=2mm]}] (-1,-1)--(0.4,0);

        \draw[thick, -{Latex[length=2mm]}] (0.5,0)--(2,-1.35);
        \draw[thick, fill=green](1.65,-1.5)--(2.35,-1.5)--(2.35,-1.35)--(1.65,-1.35)--(1.65,-1.5);
        \draw[thick, -{Latex[length=2mm]}] (1.2,-2.2)--(2,-1.5);
        \draw[thick](0.6,-2.0)--(0.6,-2.3)--(1.2,-2.3)--(1.2,-2.0)--(0.6,-2.0);
        \draw[thick, -{Latex[length=2mm]}] (2.2,-1.5)--(3.2,-2.2);
         \draw[thick, fill=yellow](2.95,-2.2)--(3.45,-2.2)--(3.45,-2.3)--(2.95,-2.3)--(2.95,-2.2);
         \draw[thick,-{Latex[length=2mm]} ] (2.7,-2.7)--(3.2,-2.3);
          \draw[thick,-{Latex[length=2mm]} ] (3.2,-2.20)--(3.5,-1.95);
           \draw[thick,-{Latex[length=2mm]} ] (3.2,-2.3)--(3.5,-2.5);
           \draw [thick, fill=red] (3.5, -1.95) -- (3.4,-1.86)..controls (3.5,-1.8) and(3.7,-1.8) ..(3.6,-2.04)--(3.5,-1.95);
          \draw [thick, fill=red] (3.5, -2.5) -- (3.6,-2.39)..controls (3.7,-2.65)and (3.5,-2.65) ..(3.4,-2.61)--(3.5,-2.5);
           \filldraw[thick, color=black, fill=white] (4,-2.22) circle (0.1);
           \draw[thick] (3.6,-1.85) to [bend left=90] (4,-2.1);
           \draw[thick] (3.6,-2.6) to [bend right=90] (4,-2.3);
           \draw[thick, -{Latex[length=2mm]}] (2,-1.35)--(3.5,0);
           
           \draw[thick, -{Latex[length=2mm]}] (0.5,0.15)--(2,1.5);
           \draw[thick, fill=green] (1.65,1.65)--(2.35,1.65)--(2.35,1.5)--(1.65,1.5)--(1.65,1.65);
           \draw[thick, -{Latex[length=2mm]}] (1.2,2.25)--(2,1.65);
          \draw[thick](0.6,2.10)--(0.6,2.4)--(1.2,2.4)--(1.2,2.10)--(0.6,2.10);
        \draw[thick, -{Latex[length=2mm]}] (2.1,1.65)--(3.2,2.5);
         \draw[thick, fill=yellow](2.95,2.5)--(3.45,2.5)--(3.45,2.6)--(2.95,2.6)--(2.95,2.5);
         \draw[thick,-{Latex[length=2mm]} ] (2.8,2.9)--(3.2,2.6);
           \draw[thick,-{Latex[length=2mm]} ] (3.2,2.5)--(3.5,2.3);
           \draw[thick,-{Latex[length=2mm]} ] (3.2,2.6)--(3.55,2.85); 
           \draw [thick, fill=red] (3.5, 2.28) -- (3.4,2.19)..controls (3.5,2.13) and(3.7,2.13) ..(3.6,2.37)--(3.5,2.28);
           \draw [thick, fill=red] (3.52, 2.85) -- (3.62,2.74)..controls (3.72,3)and (3.52,3) ..(3.42,2.94)--(3.52,2.85);
           \filldraw[thick, color=black, fill=white] (4,2.6) circle (0.1);
           \draw[thick](3.6,2.25) to [bend right=90] (4,2.5);
           \draw[thick] (3.6,2.95) to [bend left=90] (4,2.7);
            \draw[thick, -{Latex[length=2mm]}] (2.1,1.5)--(3.5,0.15);
          \draw[thick, fill=green] (3.15,0)--(3.85,0)--(3.85,0.15)--(3.15,0.15)--(3.15,0);
          \draw[thick, -{Latex[length=2mm]}] (3.5,0.15)--(5,0.85);
          \draw[thick, -{Latex[length=2mm]}] (3.5,0)--(5,-0.85);
         
          \draw[] (-1.6,1)--(-1.6,1) node []{\scriptsize $(X_{\mathrm M},Y_{\mathrm M})$};
          \draw[] (-1.5,-1)--(-1.5,-1) node []{\scriptsize $(X_{R},Y_{R})$};
          \draw[] (-0.3,0.08)--(-0.3,0.08) node []{\scriptsize $\text{BS1}$};
          \draw[] (1.37,-1.42)--(1.37,-1.42) node []{\scriptsize $\text{BS4}$};
          \draw[] (1.35,1.6)--(1.35,1.6) node []{\scriptsize $\text{BS3}$};
          \draw[] (0.9,2.25)--(0.9,2.25) node []{\scriptsize $\text{NL}$};
           \draw[] (0.9,-2.15)--(0.9,-2.15) node []{\scriptsize $\text{NL}$};
            \draw[] (2.5,-2.75)--(2.5,-2.75) node []{\scriptsize $\text{LO}$};
            \draw[] (2.6,2.95)--(2.6,2.95) node []{\scriptsize $\text{LO}$};
            \draw[] (4,-2.25)--(4,-2.25) node []{\scriptsize -};
            \draw[] (4,2.57)--(4,2.57) node []{\scriptsize -};
             \draw[] (2.85,0.08)--(2.85,0.08) node []{\scriptsize $\text{BS2}$};
             \draw[] (5.8,0.9)--(5.8,0.9) node []{\scriptsize $(X^{\text{(out)}}_{\mathrm M},Y^{\text{(out)}}_{\mathrm M})$};
          \draw[] (5.77,-0.9)--(5.77,-0.9) node []{\scriptsize $(X^{\text{(out)}}_{R},Y^{\text{(out)}}_{R})$};
          \draw[] (1.1,-3.2)--(1.1,-3.2) node []{\scriptsize \text{Squeezer A}};
          \draw[] (1.1,3.4)--(1.1,3.4) node []{\scriptsize \text{Squeezer B}};
          \draw[thick, dotted] (-2.4, 1.5)--(-0.8,1.5)--(-0.8,-1.2)--(-2.4,-1.2)--(-2.4,1.5);
          \draw[] (-2,1.3)--(-2,1.3) node []{\scriptsize \text{Input}};
           \draw[thick, dotted] (6.6, 1.5)--(4.7,1.5)--(4.7,-1.2)--(6.6,-1.2)--(6.6,1.5);
          \draw[] (5.15,1.3)--(5.15,1.3) node []{\scriptsize \text{Output}};
          \draw[](1.9,-0.6)--(1.9,-0.6) node []{\scriptsize $(X^{1}_{\mathrm M},Y^{1}_{\mathrm M})$};
           \draw[](1.9,0.75)--(1.9,0.75) node []{\scriptsize $(X^{1}_{R},Y^{1}_{R})$};
           \draw[](3.53,-0.6)--(3.53,-0.6) node []{\scriptsize $(X^{2}_{\mathrm M},Y^{2}_{\mathrm M})$};
           \draw[](3.5,0.9)--(3.5,0.9) node []{\scriptsize $(X^{2}_{R},Y^{2}_{R})$};
           \draw[] (3.28,-1.55)--(3.28,-1.55) node []{\scriptsize $(\tilde{X}^{\text{(HD)}}_{A},\tilde{Y}^{\text{(HD)}}_{A})$};
            \draw[] (1.15,-2.6)--(1.15,-2.6) node []{\scriptsize $(X^{\text{(HD)}}_{A},Y^{\text{(HD)}}_{A})$};
           \draw[] (3.28,1.85)--(3.28,1.85) node []{\scriptsize$(\Tilde{X}^{\text{(HD)}} _{B},\tilde{Y}^{\text{(HD)}}_{B})$};
            \draw[] (1.15,2.65)--(1.15,2.65) node []{\scriptsize $(X^{\text{(HD)}}_{B},Y^{\text{(HD)}}_{B})$};
            \draw[thick, dotted] (0.3,3.6)--(4.5,3.6)--(4.5,1.2)--(0.3,1.2)--(0.3,3.6);
             \draw[thick, dotted] (0.3,-3.4)--(4.5,-3.4)--(4.5,-1)--(0.3,-1)--(0.3,-3.4);
    \end{tikzpicture}
    \caption{ schematic of of the CNOT receiver based on \cite{yoshikawa2008demonstration}. In both squeezer circuits, the NL module stands for a non linear parametric element that can be an optical parametric oscillator as in the optical domain, or  a Josephson parametric amplifier as in the microwave domain. As described in appendix \ref{sec:Analysis}, both local oscillator fields (LO) are mixed with the outputs of the 'g' beamsplitters on another balanced beamsplitters. Squeezer A is momentum squeezed, whereas squeezer B is position squeezed. }
    \label{fig:model}
\end{figure}
In this section we present an implementation of the CNOT gate receiver in the main text. This model will also serve as a practical representation of the receiver's internal noise, which will eventually play a role when calculating the receiver's overall noise variance.  Following the experimental implementation presented in \cite{yoshikawa2008demonstration}, and the theoretical study in \cite{filip2005measurement}, the first beamsplitter (BS1) in Fig. (\ref{fig:model}) is described by  
\begin{align}
    \begin{pmatrix}
        \sqrt{\frac{g}{1+g}} & \sqrt{\frac{1}{1+g}} \\ 
        -\sqrt{\frac{1}{1+g}} & \sqrt{\frac{g}{1+g}}
    \end{pmatrix}
\end{align}
The signal-idler quadratures transform as 
\begin{align}
    X^{1}_{\mathrm M} &= \sqrt{\frac{g}{1+g}}  X_{\mathrm M} + \sqrt{\frac{1}{1+g}}X_{R} \nonumber \\ 
     X^{1}_{R} &= \sqrt{\frac{g}{1+g}}  X_{R} - \sqrt{\frac{1}{1+g}}X_{\mathrm M} \nonumber \\
      Y^{1}_{\mathrm M} &= \sqrt{\frac{g}{1+g}}  Y_{\mathrm M} + \sqrt{\frac{1}{1+g}} Y_{R}\nonumber \\
      Y^{1}_{R} &= \sqrt{\frac{g}{1+g}} Y_{R} - \sqrt{\frac{1}{1+g}} Y_{\mathrm M}\nonumber \\
\end{align}
Then each output of the first beamsplitter is mixed on another beamsplitter of transmissivity $1-g$ with the outputs of two single mode squeezers, such that squeezer A is momentum squeezed, i.e., $X^{\text{(HD)}}_{A} e^{r_{A}}$, $Y^{\text{(HD)}}_{A} e^{-r_{A}}$, on the other hand squeezer B is position squeezed, i.e., $X^{\text{(HD)}}_{B} e^{-r_{B}}$, $Y^{\text{(HD)}}_{B} e^{r_{B}}$. The beamsplitters are denoted by BS4, BS3 respectively. For ease of readability, we omit the exponential factors from the squeezed mode in the upcoming derivation, then add them back in the last step. 
\begin{align}
    \begin{pmatrix}
        \sqrt{1-g} & \sqrt{g} \\
        \sqrt{g} & -\sqrt{1-g}
    \end{pmatrix} .
\end{align}
Thus the modes transform as,
\begin{align}
    X^{(2)}_{\mathrm M} &= \frac{g}{\sqrt{1+g}} X_{\mathrm M} + \sqrt{\frac{g}{1+g}}X_{R} + \sqrt{1-g} X^{\text{(HD)}}_{A} \nonumber \\ 
    \Tilde{X}^{\text{(HD)}}_{A} &= \sqrt{g} X^{\text{(HD)}}_{A} - \sqrt{\frac{g(1-g)}{1+g}} X_{\mathrm M} - \sqrt{\frac{1-g}{1+g}}X_{R} \nonumber \\
   X^{(2)}_{R} &= \frac{g}{\sqrt{1+g}}  X_{R} - \sqrt{\frac{g}{1+g}}X_{\rm M} + \sqrt{1-g} X^{\text{(HD)}}_{B} \nonumber \\ 
   \tilde{X}^{\text{(HD)}}_{B} &= \sqrt{g}X^{\text{(HD)}}_{B} -\sqrt{\frac{(1-g)g}{1+g}}X_{R}+ \sqrt{\frac{1-g}{1+g}}X_{\mathrm M}
   \label{eq:homPractical}
\end{align}
Similarly,
\begin{align}
Y^{(2)}_{\mathrm M} &= \frac{g}{\sqrt{1+g}} Y_{\mathrm M} + \sqrt{\frac{g}{1+g}}Y_{R} + \sqrt{1-g} Y^{\text{(HD)}}_{A} \nonumber \\ 
    \Tilde{Y}^{\text{(HD)}}_{A} &= \sqrt{g} Y^{\text{(HD)}}_{A} - \sqrt{\frac{g(1-g)}{1+g}} Y_{\mathrm M} - \sqrt{\frac{1-g}{1+g}}Y_{R} \nonumber \\
   Y^{(2)}_{R} &= \frac{g}{\sqrt{1+g}}  Y_{R} - \sqrt{\frac{g}{1+g}}Y_{\mathrm M} + \sqrt{1-g} Y^{\text{(HD)}}_{B} \nonumber \\ 
   \tilde{Y}^{\text{(HD)}}_{B} &= \sqrt{g}Y^{\text{(HD)}}_{B} -\sqrt{\frac{(1-g)g}{1+g}}Y_{R}+ \sqrt{\frac{1-g}{1+g}}Y_{\mathrm M}
   \label{eq:homPractical1}
\end{align}
Finally the modes labeled by the superscript '(HD)' are homodyned with a local oscillator field (LO), whereas the other modes are directed towards a final beamsplitter (BS2) of transmissivity $\frac{1}{1+g}$ 
\begin{align}
    \begin{pmatrix}
        \sqrt{\frac{1}{1+g}} & \sqrt{\frac{g}{1+g}}  \\ 
        -\sqrt{\frac{g}{1+g}} & \sqrt{\frac{1}{1+g}}
    \end{pmatrix}
\end{align}
Let us consider first the position quadratures and see how they evolve,
\begin{align}
    X^{\text{(out)}}_{R} &=   \sqrt{\frac{1}{1+g}}X^{(2)}_{R} + \sqrt{\frac{g}{1+g}} X^{(2)}_{\mathrm M} \nonumber \\
    &= \Big( \frac{2g}{1+g}\Big)X_{R} - \Big( \frac{\sqrt{g}(1-g)}{1+g}\Big)X_{\mathrm M} \nonumber \\
    &+ \sqrt{\frac{1-g}{1+g}}X^{\text{(HD)}}_{B} + \sqrt{\frac{g(1-g)}{1+g}}X^{\text{(HD)}}_{A} \nonumber \\
    X^{\text{(out)}}_{\mathrm M} &= \sqrt{\frac{1}{1+g}} X^{(2)}_{\mathrm M}- \sqrt{\frac{g}{1+g}}X^{(2)}_{R} \nonumber \\
    &=\Big( \frac{2g}{1+g}\Big)X_{\mathrm M} + \Big( \frac{\sqrt{g}(1-g)}{1+g}\Big)X_{R} \nonumber \\ 
    &+ \sqrt{\frac{1-g}{1+g}}X^{\text{(HD)}}_{A} - \sqrt{\frac{g(1-g)}{1+g}}X^{\text{(HD)}}_{B},
    \label{eq:ScalePractical1}
\end{align}
Suppose now that the mode $\Tilde{X}^{\text{(HD)}}_{A}$ in Eq. (\ref{eq:homPractical}) was homodyned with efficiency $\gamma$, that is, $\sqrt{\gamma} \Tilde{X}^{\text{(HD)}}_{A} - \sqrt{1-\gamma}X_{V}$, where $X_{V}$ is a vacuum position quadrature, then after being re-scaled appropriately is utilised to perform the following post-correction operation in order to eliminate the anti-squeezed position quadrature $X^{\text{(HD)}}_{A}$,
\begin{align}
    X^{\text{(out)}}_{R} & \rightarrow X^{\text{(out)}}_{R} - \sqrt{\frac{1-g}{\gamma (1+g)}}\Tilde{X}^{\text{(HD)}}_{A} \nonumber \\ 
    &\rightarrow X^{\text{(out)}}_{R}- \sqrt{\frac{g(1-g)}{1+g}} X^{\text{(HD)}}_{A}+ \frac{1-g}{1+g}X_{R} \nonumber \\ 
    &+\frac{\sqrt{g}(1-g)}{(1+g)} X_{\mathrm M} + \sqrt{\frac{(1-\gamma)(1-g)}{\gamma g(1+g)}}X_{V}, \nonumber 
    \end{align}
    \begin{align}
    X^{\text{(out)}}_{R} &= X_{R} + \sqrt{\frac{1-g}{1+g}}X^{\text{(HD)}}_{B}+ \sqrt{\frac{(1-\gamma)(1-g)}{\gamma g(1+g)}}X_{V}, \nonumber \\
\label{eq:ScalePractical2}
\end{align}

We follow a similar approach to derive the expression of $X^{\text{(out)}}_{\mathrm M}$, where a different appropriate scaling of  $\Tilde{X}^{\text{(HD)}}_{A}$ is assumed as follows;
\begin{align}
    X^{\text{(out)}}_{\mathrm M} &\rightarrow X^{\text{(out)}}_{\mathrm M} - \sqrt{\frac{(1-g)}{\gamma g(1+g)}} \tilde{X}^{\text{(HD)}}_{A} \nonumber \\ 
    &\rightarrow X^{\text{(out)}}_{\mathrm M} -\sqrt{\frac{(1-g)}{(1+g)}}X^{\text{(HD)}}_{A}+\frac{1-g}{1+g} X_{\mathrm M}  \nonumber \\ 
    &+\frac{(1-g)}{\sqrt{g}(1+g)} X_ {R}+\sqrt{\frac{(1-\gamma)(1-g)}{\gamma (1+g)}}X_{V} \nonumber \\ 
    &= X_{\mathrm M}+ \Big(\frac{1-g}{1+g}(\sqrt{g}+\frac{1}{\sqrt{g}}) \Big)X_{R}-\sqrt{\frac{g(1-g)}{1+g}}X^{\text{(HD)}}_{B}\nonumber \\ &+\sqrt{\frac{(1-\gamma)(1-g)}{\gamma g(1+g)}}X_{V} \nonumber \\
    &=X_{\mathrm M}+ \Big(\frac{1-g}{\sqrt{g}} \Big)X_{R}-\sqrt{\frac{g(1-g)}{1+g}}X^{\text{(HD)}}_{B}\nonumber \\ &+\sqrt{\frac{(1-\gamma)(1-g)}{\gamma g(1+g)}}X_{V}
    \label{eq:ScalePractical3}
\end{align}
where $G=\frac{1-g}{\sqrt{g}}$.

Focusing now on the momentum quadratures we follow a similar derivation to that in Eqs. (\ref{eq:ScalePractical1}-\ref{eq:ScalePractical3}), 
\begin{align}
    Y^{\text{(out)}}_{R} &=   \sqrt{\frac{1}{1+g}}Y^{(2)}_{R} + \sqrt{\frac{g}{1+g}} Y^{(2)}_{\mathrm M} \nonumber \\
    &= \Big( \frac{2g}{1+g}\Big)Y_{R} - \Big( \frac{\sqrt{g}(1-g)}{1+g}\Big)Y_{\mathrm M} + \sqrt{\frac{1-g}{1+g}}Y^{\text{(HD)}}_{B} \nonumber \\ 
    &+ \sqrt{\frac{g(1-g)}{1+g}}Y^{\text{(HD)}}_{A} \nonumber 
    \end{align}
    \begin{align}
    Y^{\text{(out)}}_{\mathrm M} &= \sqrt{\frac{1}{1+g}} Y^{(2)}_{\mathrm M}- \sqrt{\frac{g}{1+g}}Y^{(2)}_{R} \nonumber \\
    &=\Big( \frac{2g}{1+g}\Big)Y_{\mathrm M} + \Big( \frac{\sqrt{g}(1-g)}{1+g}\Big)Y_{R} + \sqrt{\frac{1-g}{1+g}}Y^{\text{(HD)}}_{A} \nonumber \\ 
    &- \sqrt{\frac{g(1-g)}{1+g}}Y^{\text{(HD)}}_{B}, 
\end{align}
Then similarly, we assume that $\tilde{Y}^{\text{(HD)}}_{B}$ in Eq. (\ref{eq:homPractical1}) was homodyned with efficiency $\gamma$, and used to perform the following post-correction operation after proper re-scaling in order to eliminate the anti-squeezed momentum quadrature $Y^{\text{(HD)}}_{B}$
\begin{align}
    Y^{\text{(out)}}_{R} &\rightarrow Y^{\text{(out)}}_{R} - \sqrt{\frac{1-g}{\gamma g(1+g)}}\Tilde{Y}^{\text{(HD)}}_{B} \nonumber \\ 
    &\rightarrow Y^{\text{(out)}}_{R}- \sqrt{\frac{1-g}{1+g}} Y^{\text{(HD)}}_{B}+ \frac{1-g}{1+g}Y_{R} \nonumber \\ 
    &-\frac{(1-g)}{\sqrt{g}(1+g)} Y_{\mathrm M} + \sqrt{\frac{(1-\gamma)(1-g)}{\gamma g(1+g)}}Y_{V} \nonumber \\
    &= Y_{R} - \Big(\frac{1-g}{1+g}(\sqrt{g}+\frac{1}{\sqrt{g}}) \Big)Y_{\mathrm M}+ \sqrt{\frac{g(1-g)}{1+g}}Y^{\text{(HD)}}_{A} \nonumber \\ &+\sqrt{\frac{(1-\gamma)(1-g)}{\gamma g(1+g)}}Y_{V} \nonumber \\
     &= Y_{R} - G Y_{\mathrm M}+ \sqrt{\frac{g(1-g)}{1+g}}Y^{\text{(HD)}}_{A} \nonumber \\ &+\sqrt{\frac{(1-\gamma)(1-g)}{\gamma g(1+g)}}Y_{V} \nonumber \\
\end{align}
Similarly, 
\begin{align}
    Y^{\text{(out)}}_{\mathrm M} &\rightarrow Y^{\text{(out)}}_{\mathrm M}+\sqrt{\frac{(1-g)}{\gamma (1+g)}}\tilde{Y}^{\text{(HD)}}_{B} \nonumber \\
    &\rightarrow Y^{\text{(out)}}_{\mathrm M}+\frac{1-g}{1+g}Y_{\mathrm M}-\frac{\sqrt{g}(1-g)}{(1+g)} Y_{R} \nonumber \\ &+ \sqrt{\frac{(1-\gamma)(1-g)}{\gamma (1+g)}}Y_{V}, \nonumber\\
   Y^{\text{(out)}}_{\mathrm M} &=Y_{\mathrm M} - \sqrt{\frac{g(1-g)}{1+g}}Y^{\text{(HD)}}_{A} + \sqrt{\frac{(1-\gamma)(1-g)}{\gamma g(1+g)}}Y_{V}
\end{align}
Therefore the four receiver's output can be written as 
\begin{align}
   X^{\text{(out)}}_{R} &= X_{R} + \sqrt{\frac{1-g}{1+g}}X^{\text{(HD)}}_{B}e^{-r_{B}}+ \sqrt{\frac{(1-\gamma)(1-g)}{\gamma g(1+g)}}X_{V}, \nonumber \\
  X^{\text{(out)}}_{\mathrm M} &=X_{\mathrm M}+ GX_{R}-\sqrt{\frac{g(1-g)}{1+g}}X^{\text{(HD)}}_{B}e^{-r_{B}}\nonumber \\ &+\sqrt{\frac{(1-\gamma)(1-g)}{\gamma g(1+g)}}X_{V} \nonumber \\
  Y^{\text{(out)}}_{R}&= Y_{R} - G Y_{\mathrm M}+ \sqrt{\frac{g(1-g)}{1+g}}Y^{\text{(HD)}}_{A}e^{-r_{A}} \nonumber \\ &+\sqrt{\frac{(1-\gamma)(1-g)}{\gamma g(1+g)}}Y_{V} \nonumber \\
  Y^{\text{(out)}}_{\mathrm M} &=Y_{\mathrm M} - \sqrt{\frac{g(1-g)}{1+g}}Y^{\text{(HD)}}_{A}e^{-r_{A}} + \sqrt{\frac{(1-\gamma)(1-g)}{\gamma g(1+g)}}Y_{V}
\end{align}
It can be seen from the above equation that the ideal transformation in Eq.(\ref{eq:CNOTRXOUT}) can be retrieved in the limit of large squeezing parameters $r_{A}$, $r_{B}$ and unity homodyne detection efficiency $\gamma$.

We now consider the effect of finite squeezing and inefficient homodyne detection on the overall number of added noise photons. From the previous equation the total noise power can be calculated as
\begin{align}
    \big \langle [X^{\text{(out)}}_{R}]^{2} \big \rangle &= \frac{\langle X^{2}_{B} \rangle}{2} +\frac{1-g}{2(1+g)} \big \langle [X^{\text{(HD)}}_{B}]^{2} \big \rangle +\frac{\langle X^{2}_{V}\rangle}{2} \nonumber \\ 
    &+ \frac{(1-\gamma)(1-g)}{2\gamma g (1+g)} \langle X^{2}_{V} \rangle \nonumber \\ 
    \big \langle [X^{\text{(out)}}_{\mathrm M}]^{2} \big \rangle &= \frac{\langle X^{2}_{\mathrm M} \rangle}{2} + \frac{g(1-g)}{2(1+g)} \big \langle [X^{\text{(HD)}}_{B}]^{2} \big \rangle +\frac{G^{2}\langle X_{B}\rangle}{2} \nonumber \\ 
    &+ \frac{(1-\gamma)(1-g)}{2\gamma  g(1+g)} \langle Y^{2}_{V} \rangle + \frac{\langle Y^{2}_{V}\rangle}{2}\nonumber  
    \end{align}
    \begin{align}
    \big \langle [Y^{\text{(out)}}_{R}]^{2} \big \rangle &= \frac{\langle Y^{2}_{B} \rangle}{2} + \frac{g(1-g)}{2(1+g)} \big \langle [Y^{\text{(HD)}}_{A}]^{2} \big \rangle + \frac{\langle X^{2}_{V} \rangle}{2}\nonumber \\ 
    &+ \frac{(1-\gamma)(1-g)}{2\gamma  g(1+g)} \langle X^{2}_{V} \rangle + \frac{G^{2}Y_{\mathrm M}}{2}\nonumber \\ 
     \big \langle [Y^{\text{(out)}}_{\mathrm M}]^{2} \big \rangle &= \frac{\langle Y^{2}_{\mathrm M} \rangle}{2} + \frac{\langle Y^{2}_{V} \rangle}{2}+\frac{g(1-g)}{2(1+g)} \big \langle [Y^{\text{(HD)}}_{A}]^{2} \big \rangle \nonumber \\ &+ \frac{(1-\gamma)(1-g)}{2\gamma g (1+g)} \langle Y^{2}_{V} \rangle \nonumber \\ 
    \label{eq:VarPractical}
\end{align}
The homodyne inefficiency and the finite squeezing of the utilized squeezer circuits enter the picture as extra added noise, and we have added the 3dB loss penalty due to measuring non commuting quadratures. By recalling that the value of of the beamsplitter parameter is $0<g<1$, and the low brightness regime, it can be seen that the bath noise power dominates and the overall noise power is the expression derived earlier in Eq. (\ref{eq:ADDNOISE}). 

For practical considerations, the following experimental parameters can be assumed for physical implementation of the CNOT receiver. A realistic squeezing that can be achieved in a laboratory is approximately equal to $\approx -3 \text{dB}$, that is, $e^{-2r} \approx 0.5$, such that $r = \ln{2} / 2$ \cite{esposito2022observation, qiu2023broadband}. It is also possible to achieve up to $-6 \text{dB}$ experimentally \cite{fedorov2016displacement}. As for practical gain values when a JPA is utilised as a squeezing resource, the optimal gain values are approximately $\approx 15 \pm 3\text{dB}$. In this regime the JPA remains quantum limited, i.e. only adds half a quantum of noise. Finally, in the optical domain the homodyne detector's efficiency is  approximately $\gamma \approx 0.97$ \cite{furusawa2011quantum}. Recently graphene-based microwave bolometers \cite{walsh2017graphene, lee2020graphene, kokkoniemi2020bolometer} have enjoyed similar successes, and thus in either cases it is pretty reasonable to assume ideal operation. Thus by adding the squeezing and vaccuum noise contributions in Eq. (\ref{eq:VarPractical}), we estimate that the device internal noise adds approximately $\approx 2$ noise photons.

\end{appendices}
\section*{Acknowledgment}
The authors are grateful to Maximilian Reichert, Roberto Di Candia, Robert Jonsson, and Stefano Pirandola for discussions at various stages of this work.
\EOD
\providecommand{\noopsort}[1]{}\providecommand{\singleletter}[1]{#1}%

\begin{IEEEbiography}{Hany Khalifa} (hany.khalifa@aalto.fi) received his M.Sc. from Ain Shams University, Faculty of Engineering, Cairo, Egypt in 2018. Currently he is finalizing his doctoral studies at Aalto University. His research interests include quantum optics, quantum information theory and quantum communications. 
\end{IEEEbiography}
\begin{IEEEbiography}{Kirill Petrovnin} (kirill.petrovnin@aalto.fi) received his PhD (2019) in Radiophysics and Optics from Kazan Federal University, Russia. He is currently a Postdoctoral researcher in Department of Applied Physics at Aalto University, Finland. His research mainly focused on continuous variable entanglement, sensing and information protocols, achievable with superconducting quantum parametric devices, which operate in microwave frequencies.

\end{IEEEbiography}
\begin{IEEEbiography}{Riku J{\"a}ntti} [SM'07] (riku.jantti@aalto.fi) received his M.Sc. degree (with distinction) in electrical engineering in 1997 and his D.Sc. degree (with distinction) in automation and systems technology in 2001, both from Helsinki University of Technology (now Aalto), Finland. He is a full professor of communications engineering and the head of the Department of Communications and Networking at Aalto University School of Electrical Engineering, Finland. He is an associate editor of IEEE Transactions on Vehicular Technology. He is also IEEE VTS Distinguished Lecturer (Class 2016). His research interests include radio resource control and optimization for machine-type communications, cloud-based radio access networks, spectrum and coexistence management, and quantum communications.
\end{IEEEbiography}
\begin{IEEEbiography}{G.S. Paraoanu }  (sorin.paraoanu@aalto.fi)  did his B.Sc. and M.Sc. in physics at the University of Bucharest, Romania, followed by a Ph.D. in theoretical physics (2001) from the University of Illinois at Urbana-Champaign, U.S.A. He went then for a postdoc at the University of Jyväskylä, Finland - followed by several academic positions (Marie Curie scholarship, senior research scientist, Academy of Finland Research Fellow). He also had visiting positions in Austria and Switzerland. He is presently at Aalto University in Finland (Senior University Lecturer) where he leads a group specialized in superconducting quantum circuits.
\end{IEEEbiography}
\EOD
\end{document}